\documentclass[10pt,aps,prd,nofootinbib,superscriptaddress,twocolumn]{revtex4}
\usepackage{tikz}
\usepackage[utf8]{inputenc}
\usepackage{color}
\usepackage{graphicx}
\usepackage{float}
\usepackage{amsmath}
\usepackage{amssymb}
\usepackage{bm}
\usepackage{acronym}
\usepackage{ifthen}
\usepackage{blindtext}
\usepackage[normalem]{ulem}
\usepackage{hyperref}
\usepackage{etoolbox}
\usepackage{fancyhdr}
\usepackage{xspace}
\usepackage{textcomp}
\usepackage{multirow}
\usepackage[caption=false]{subfig}
\usepackage{lineno}
\usepackage{tabularx}
\usepackage{float}
\hypersetup{
    colorlinks=true,
    linkcolor=blue,
    filecolor=magenta,      
    urlcolor=black,
		citecolor=red,
		}
\begin{document}
	
	\title{Energy Extraction from Rotating Black Hole with Quintessential Energy through the Penrose Process}
	\author{K. Q. Abbasi}
	\email{kamranqadir@numl.edu.pk}
	\affiliation{Department of Mathematics\\
		Faculty of Engineering and Computing (FE\&C)\\
		National University of Modern Languages (NUML),\\
		Sector H-9, Islamabad, Pakistan.\\}
	\author{F. L. Carneiro}
	\email{fernandolessa45@gmail.com}
	\affiliation{Universidade Federal do Norte do Tocantins, \\
		77824-838, Aragua\'ina, TO, Brazil.}
	\author{M. Z. A. Moughal}
	\email{moughalzubair@gmail.com}
	\affiliation{Department of Basic Sciences and Humanities,\\
		College of Electrical and Mechanical Engineering,\\
		National University of Science and Technology,\\
		Islamabad, Pakistan}
	\date{\today}
	\begin{abstract}
		We investigate the geometry, dynamics, and collision mechanisms in the ergoregion of Kerr-Newman-AdS black hole influenced by quintessential energy. Particle splittings within the ergoregion are analyzed, demonstrating their role in energy extraction via the Penrose process. Increased spin elongates the ergosphere, while higher quintessential parameters expand static limits and distort photon regions. Prograde orbits benefit from reduced energy and angular momentum due to frame-dragging, whereas retrograde orbits require higher energy. Quintessential energy weakens the gravitational pull, shifts stable orbit radii, and enhances orbital chaos, as indicated by Lyapunov exponents. The Penrose process demonstrates efficiencies ranging from 5\% to 35\%, with peak efficiency achieved at high spin, but diminishing with increased charge or quintessential energy due to reduced frame-dragging. We derive the exprssion for irreducible mass and discuss its dependence on cosmological and quintessence parameters, revealing their role in limiting extractable energy.
	\end{abstract}
	
	\pacs{04.20.-q, 04.20.Cv} 
	
	\maketitle
	\section{Introduction}
	
	The study of black holes (BHs) in cosmological settings, particularly rotating BHs described by the Kerr-Newman-AdS (KNAdS) solution, has attracted significant attention due to their intricate structure and astrophysical relevance. The KNAdS solution \cite{Newman1965} generalizes the Kerr-Newman metric by incorporating charge, spin, and a cosmological constant, offering a comprehensive framework for exploring BHs in a universe dominated by dark energy. Of particular interest is the influence of cosmological fields like quintessence—a dynamic form of dark energy—on BH physics. Quintessence, characterized by a time-dependent equation of state parameter (\(\omega\)), contrasts with the static nature of the cosmological constant and significantly alters the surrounding spacetime \cite{Kiselev2003}. These investigations deepen our understanding of how BHs interact with an evolving universe shaped by dark energy.
	
	Dark energy, which drives the accelerated expansion of the universe, constitutes approximately 68\% of its total energy density, while dark matter and baryonic matter account for the remaining composition \cite{Farnes2018}. Dark energy is commonly modeled as either a small positive cosmological constant (\(\Lambda > 0\))—with an approximate value of \(\Lambda \sim 10^{-52} \, \mathrm{m^{-2}}\)—or as a dynamic scalar field like quintessence. The cosmological constant is uniform throughout space and exerts negative pressure, fundamentally shaping spacetime. Its effects include altering the asymptotic properties of BHs and singularities, resulting in geometries that are asymptotically de Sitter or anti-de Sitter. By linking the cosmological constant and dynamic fields like quintessence to the behavior of BHs, these studies provide critical insights into the interplay between gravitational phenomena and the large-scale evolution of the universe.
	
	The interaction between BH geometry and quintessence has been a topic of significant interest. The presence of quintessence energy can significantly modify key properties of BHs, such as the ergosphere, static limit, and event horizons, leading to altered energy dynamics and orbital stability around BHs \cite{Banados2009}. In particular, quintessence alters the spacetime structure of KNAdS BH by modifying the geometry of the static limit and ergosphere—the regions where particle motion is constrained. These modifications are influenced by the BH's spin, mass, and the quintessence parameter, all contributing to a dynamic and evolving gravitational environment \cite{Pati1995}. 
	
	Studies have shown that the spin of a BH affects the size and shape of the ergosphere, with higher spins leading to an extended and more oblate ergosphere, which can influence particle dynamics and energy extraction processes such as the Penrose process \cite{Penrose1969,Thorne1974}. In addition to spin, the charge-to-mass ratio (\(Q/M\)) also plays a significant role in determining the energy available for extraction, as it influences the frame-dragging effect that is central to these processes \cite{Gibbons2004}.
	
	Recent developments in BH thermodynamics and accretion theory  \cite{AdvancesBH2023} have emphasized the importance of understanding energy extraction mechanisms, such as the Penrose process \cite{Penrose1971}, which is particularly sensitive to the BH's geometry. The efficiency of energy extraction is influenced by the interplay between the BHs spin, charge, quintessence parameter, and other environmental factors, including the radial distance from the BH \cite{Berti2009}.
	
	The study of prograde and retrograde orbits provides crucial insights into the dynamics of test particles in the curved spacetime of KNAdS BH. Prograde motion, which aligns with the BHs rotation, benefits from the frame-dragging effect, requiring less energy for stable orbits, while retrograde motion experiences an opposite effect, requiring higher energy \cite{NatarajanPrasanna2001}.
	
	The Lyapunov exponent, a tool for measuring orbital stability, also provides valuable insights into the chaotic behavior of particles near the BH. The Lyapunov exponent is highly sensitive to changes in BH parameters, particularly the spin and the presence of quintessence, which can either stabilize or destabilize orbits in the BH's vicinity\cite{Hawley1992}. Additionally, the efficiency of the Penrose mechanism, which allows energy extraction from the BH’s rotational energy, is intricately linked to the BH's spin and charge parameters, along with the influence of quintessence \cite{SchnittmanKrolik2009}.
	
	The KNAdS BH has been extensively studied by various authors due to its complex structure and the challenges it poses in understanding the dynamics of spacetime. Different techniques have been employed to analyze the effects of various parameters on the dynamics of particles, geodesic motion, and other related phenomena \cite{Zhang2009, Sehrish2011}. Z. Xu and J. Wang \cite{XuWang2020} provide a detailed analysis of the KNAdS BH with quintessential energy  by calculating its static limit, ergoregion, and many other aspects. However, they did not discuss the energy extraction process, even though they employed different approaches to calculate the solution and its properties. Similarly, S. Iftikhar and M. Shahzad \cite{IftikharShahzadi2021} analyze the circular motion of neutral test particles around a KN BH in the presence of quintessential dark energy. They explore the dynamics of both time-like and null geodesics, focusing on stable orbits, the photon sphere, and the static radius, while also discussing energy extraction and the Penrose process in the context of dark energy.
	
	Several papers in the literature address these topics \cite{Manna2019,Kumar2022,Kumar2022b,Zhao2010}; however, in this work, we adopt a slightly different approach to calculate the inequality for the photon region, considering a wider range of parameters and their effects. First, we define the region in which a BH exists and distinguish it from regions with naked singularities.We then calculate the angular velocity of particles in the spacetime and analyze the stability of their orbits using the effective force and Lyapunov exponents, distinguishing between prograde and retrograde motion. The chaotic behavior of test particles is also examined. Moreover, we calculate the total energy extraction from the ergoregion via the Penrose process and superradiance, providing a detailed analysis with 3D plots. We also discuss the efficiency of the energy extraction mechanism, emphasizing the conditions for negative energy states inside the ergoregion. The irreducible mass of the KNAdS BH with quintessential energy is examined, and its expression is derived in the context of energy extraction limits, demonstrating its role in setting the upper bound for rotational energy that can be extracted.
	
	The plan of the paper is as follows: In the next section, we will briefly review the properties of the KNAdS BH, with a particular focus on the horizon structure and the region defined for the set of parameters for which a BH exists, as opposed to naked singularities. In Section III, we will discuss the photon region, examining how the horizons and ergoregion are influenced by different parameters in the presence of quintessential energy. We will also calculate the angular velocity of particles, as well as their energy and angular momentum, to analyze their stability. To further study the stability, we will compute the effective force and Lyapunov exponents. In Section IV, we will discuss the Penrose process, calculate the total amount of energy extracted, and investigate the negative energy states and the efficiency of the mechanism. The analysis of the irreducible mass is also provided, along with the limitations on energy extraction through the Penrose process in this section. We will conclude with a discussion in Section V.
	
	\section{Kerr-Newman-AdS black hole solution with quintessential energy}
	The KNAdS BH solution with quintessential energy, explored by Z. Xu and J. Wang \cite{XuWang2020}, represents a solution to the Einstein–Maxwell equations, incorporating electromagnetic fields and a cosmological constant. Quintessence modifies the energy-momentum tensor, introducing a varying equation of state parameter. Characterized by mass, electric charge, angular momentum, and the cosmological constant, this solution is a more general and physically relevant description of rotating charged BHs in asymptotically de Sitter spacetimes.  The line element governing the KNAdS BH with quintessence energy is given by,
	\begin{align}
	d s^2=&\frac{\Sigma^2}{\Delta_r} d r^2+\frac{\Sigma^2}{\Delta_\theta} d \theta^2+\frac{\Delta_\theta \sin ^2 \theta}{\Sigma^2}\left(a \frac{d t}{\Xi}-\left(r^2+a^2\right) \frac{d \phi}{\Xi}\right)^2\nonumber\\
	&-\frac{\Delta_r}{\Sigma^2}\left(\frac{d t}{\Xi}-a \sin ^2 \theta \frac{d \phi}{\Xi}\right)^2,
	\label{01}
	\end{align}
	where
	\begin{align}
	& \Delta_r=r^2-2 M r+a^2+Q^2+\frac{r^2}{\ell^2}\left(r^2+a^2\right)-\alpha r^{1-3 \omega}, \nonumber\\
	& \Delta_\theta=1-\frac{a^2}{\ell^2} \cos ^2 \theta, \quad \Xi=1-\frac{a^2}{\ell^2}, \quad \Sigma^2=r^2+a^2 \cos ^2 \theta .
	\label{02}
	\end{align}
	The cosmological constant is $\Lambda=-\frac{3}{\ell^2}$ which interpret as a thermodynamic pressure $P$ by,
	\begin{align}
	P=-\frac{\Lambda}{8 \pi}=\frac{3}{8 \pi \ell^2} .
	\label{03}
	\end{align}
	For a general situation with \( -1 < \omega < -\frac{1}{3} \), (the parameter $\omega$ represents the equation of state parameter of a cosmological fluid, typically associated with  quintessence) four horizons exist: the Cauchy horizon \( r_{\text{in}} \) (or \( r_- \)), the event horizon \( r_{\text{out}} \) (or \( r_+ \)), and two cosmological horizons \( r_q \) and \( r_c \). The cosmological horizon \( r_q \) is determined by quintessential energy, while \( r_c \) is set by the cosmological constant. When the cosmological constant is zero, following the method of Toshmatov et al. \cite{toshmatov2015}, the existence of the cosmological horizon \( r_q \)requires the quintessence parameter, $\alpha$ to satisfy a specific condition. Additionally, there are two surfaces, such as the outer event horizon and the static limit surface, which meet at the poles and define a region between the horizon and static limit surface known as the ergosphere.
	
	To determine the horizon structure of these BHs, we set $\Delta_r=0$ as given in (\ref{02}). Similarly, for the stationary limit surfaces, we set $g_{tt}=0$ in \cite{XuWang2020}. For further details, refer to \cite{zhang2017horizons, astefanesei2008thermodynamics,hendi2015thermodynamics}.
	Here we have plotted different values of parameters with various combinations to determine the regions where BHs exist and where they do not. These combinations, derived from the conditions for the existence of horizons (\(\Delta_r \leq 0\)), are shown in Fig. 1. The figures illustrate how different parameters, such as the quintessence parameter, charge, and spin, influence the stability and formation of BHs, emphasizing the balance between attractive and repulsive effects in the spacetime structure.
	\begin{figure*}[t]
		\centering
		\begin{minipage}[t]{0.4\textwidth}
			\centering
			\includegraphics[width=1\linewidth]{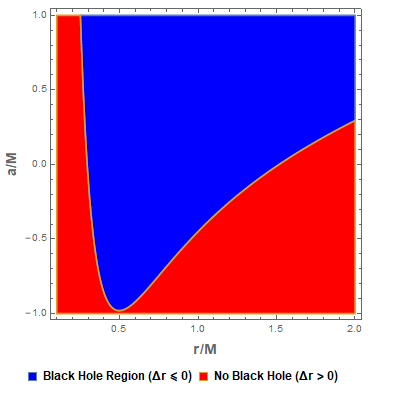}
			Fig. 1(a)
		\end{minipage}%
		\hspace{0.05\textwidth}
		\begin{minipage}[t]{0.39\textwidth}
			\centering
			\includegraphics[width=1\linewidth]{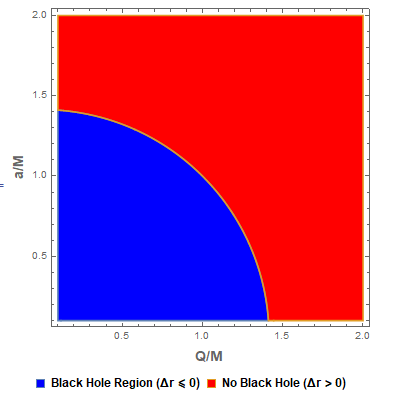}
			Fig. 1(b)
		\end{minipage}
		
		\vspace{0.5cm}
		\begin{minipage}[t]{0.45\textwidth}
			\centering
			\includegraphics[width=1.1\linewidth]{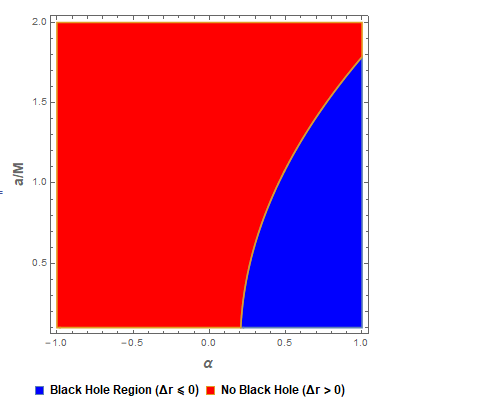}
			Fig. 1(c)
		\end{minipage}%
		\hspace{0.05\textwidth}
		\begin{minipage}[t]{0.4\textwidth}
			\centering
			\includegraphics[width=1.05\linewidth]{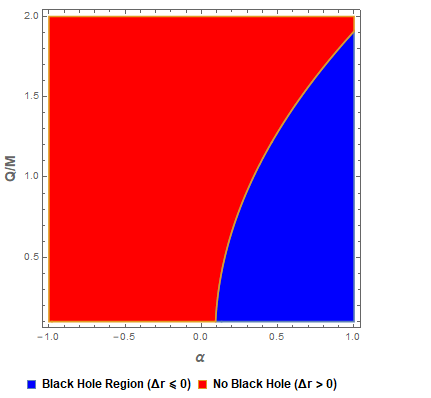}
			Fig. 1(d)
		\end{minipage}
		\captionsetup{justification=raggedright,singlelinecheck=false}
		\caption{The figures explore the effects of quintessence (\(\alpha/M\)), charge (\(Q/M\)), spin (\(a/M\)), and the radial coordinate (\(r/M\)) on the existence of horizons (\(\Delta_r \leq 0\)) in a KNAdS BH. The blue regions indicate the BH region (\(\Delta_r \leq 0\)), while the red regions represent the absence of a BH (\(\Delta_r > 0\)). Positive quintessence (\(\alpha/M > 0\)) opposes gravitational attraction, reducing the horizon region, while negative quintessence (\(\alpha/M < 0\)) enhances gravitational collapse and expands the black hole region. Increasing charge (\(Q/M\)) destabilizes the BH by introducing electric repulsion, particularly in the presence of repulsive quintessence. Similarly, higher spin (\(a/M\)) reduces the BH region due to centrifugal effects opposing gravitational collapse. The combined effects of charge and spin show that high values of both parameters significantly shrink the horizon region, emphasizing their destabilizing roles. These results highlight the delicate balance between quintessence, charge, spin, and gravitational attraction in determining  stability and horizon structure.}
		\label{Fig1}
	\end{figure*}
	\section{PHOTON REGION}
	To analyze the photon region, we focus on the motion of photons (null geodesics) in the spacetime of a KNAdS BH surrounded by quintessence. The geodesic motion is described using the Hamilton-Jacobi formalism. The Hamilton-Jacobi equation for geodesics in a spacetime with the metric $g_{\mu\nu}$
	is given by
	\begin{equation}
	\frac{\partial S}{\partial \lambda} + \frac{1}{2} g^{\mu\nu} \frac{\partial S}{\partial x^\mu} \frac{\partial S}{\partial x^\nu} = \frac{1}{2} m^2,
	\label{04}
	\end{equation}
	where \(S\) is the action, \(\lambda\) is the affine parameter, and \(x^\mu\) are the coordinates. For separability of the equation in this metric, we write
	\begin{equation}
	S = \frac{1}{2} m^2 \lambda - Et + L \phi + S_r(r) + S_\theta(\theta),
	\label{05}
	\end{equation}
	where \(E\) is the energy of the photon, \(L\) is the angular momentum, \(S_r(r)\) and \(S_\theta(\theta)\) are the radial and angular functions. For photons ($m^2=0$), the Hamilton-Jacobi equation simplifies to,
	\begin{equation}
	g^{\mu\nu} \frac{\partial S}{\partial x^\mu} \frac{\partial S}{\partial x^\nu} = 0.
	\label{06}
	\end{equation}
	Using Eq. (\ref{01}) in the Eq. (\ref{06}), we have,
	\begin{align}
	\frac{\Delta_r}{\Sigma^2} \left(\frac{\partial S_r}{\partial r}\right)^2& + \frac{\Delta_\theta}{\Sigma^2} \left(\frac{\partial S_\theta}{\partial \theta}\right)^2 + \frac{\Xi^2}{\Sigma^2 \sin^2\theta} \left(aE \sin^2\theta - L\right)^2\nonumber\\
	& - \frac{\Sigma^2}{\Delta_r} \left(E(r^2 + a^2) - aL\right)^2 = 0.
	\label{07}
	\end{align}
	We separate variables into radial and angular parts by introducing a constant \(K\) called the Carter constant. These are given by Eqs. (\ref{08}) and (\ref{09}), respectively,
	\begin{equation}
	\frac{\Delta_r}{\Sigma^2} \left(\frac{\partial S_r}{\partial r}\right)^2 - \frac{\Sigma^2}{\Delta_r} \left(E(r^2 + a^2) - aL\right)^2 + K = 0,
	\label{08}
	\end{equation}
	\begin{equation}
	\frac{\Delta_\theta}{\Sigma^2} \left(\frac{\partial S_\theta}{\partial \theta}\right)^2 + \frac{\Xi^2}{\Sigma^2 \sin^2\theta} \left(aE \sin^2\theta - L\right)^2 - K = 0.
	\label{09}
	\end{equation}
	For the photon region, photons execute circular or bound-like motion. The radial equation  Eq. (\ref{08}) must satisfy the effective potential conditions,
	\begin{equation}
	\dot{r}^2 = 0 \quad \text{and} \quad \frac{\partial (\dot{r}^2)}{\partial r} = 0.
	\label{10}
	\end{equation}
	From the radial equation, we solve the conditions:
	\begin{equation}
	\dot{r}^2 = \frac{\Delta_r}{\Sigma^2} \left(E(r^2 + a^2) - aL\right)^2 - K = 0,~~
	\frac{\partial (\dot{r}^2)}{\partial r} = 0.
	\label{11}
	\end{equation}
	This leads to a system of equations for \(r\), the radius of the photon region and constants \(L/E\) i.e, the impact parameter. For consistency, the angular equation must also be satisfied.
	
	This gives the radius \(r_p\) of the photon sphere (or photon region). Confirm that the angular motion remains bound for the constants \(E, L, K\). The quintessence term \(-\alpha r^{1-3\omega}\) alters the effective potential, shifting the photon region compared to the standard KNAdS metric. The photon region is determined by \(r_p\), which depends on the BH parameters (\(M, Q, a, \ell, \alpha, \omega\)). Analytical solutions may not be feasible for the quintessence term; numerical analysis is typically required. \(E\) is the conserved energy, \(L\) is the conserved angular momentum along the symmetry axis.
	
	The radial effective potential must satisfy Eq. (\ref{11}). These conditions determine the radius \(r_p\) of the photon sphere or the boundary of the photon region. The geodesic equations for the KNAdS BH  can be summarized as follows:
	\begin{equation}
	\dot{r} = \pm \sqrt{\frac{\Delta_r}{\Sigma^2} \left(E(r^2 + a^2) - aL\right)^2 - K},
	\label{12} 
	\end{equation}
	\begin{equation}
	\dot{\theta} = \pm \sqrt{\frac{\Delta_\theta}{\Sigma^2} \left(K - \frac{\Xi^2}{\sin^2 \theta} (aE \sin^2 \theta - L)^2\right)},
	\label{13}
	\end{equation}
	\begin{equation}
	\dot{\phi} = \frac{\Xi}{\Sigma^2 \sin^2 \theta} \left(L - aE \sin^2 \theta\right),
	\label{14}
	\end{equation}
	\begin{equation}
	\dot{t} = \frac{\Xi}{\Sigma^2} \left(E(r^2 + a^2) - aL\right).
	\label{15}
	\end{equation}
	These equations describe the photon trajectories and can be analyzed to locate the photon sphere or regions where photons are confined or execute bound orbits. The radius depends on the KNAdS BH parameters. Photons are dragged along the azimuthal direction due to the BHs rotation ($a$). The quintessence parameter $\alpha r^{1-3\omega}$ modifies the effective potential, altering the photon sphere and escape dynamics.
	
	For photons, where \(m^2 = 0\), the geodesic equations simplify to describe the motion of massless particles around the KNAdS BH.  In this case, the constant \(K\) corresponds to the impact parameter, given by \(K = \frac{L^2}{E^2}\), where \(L\) is the angular momentum and \(E\) is the energy of the photon. This leads to the radial equation becoming,
	\begin{equation}
	\dot{r} = \pm \sqrt{\frac{\Delta_r}{\Sigma^2} \left(E(r^2 + a^2) - aL\right)^2 - \frac{L^2}{E^2}}.
	\label{16}
	\end{equation}
	The angular motion for photons can be simplified as,
	\begin{equation}
	\dot{\theta} = \pm \sqrt{\frac{\Delta_\theta}{\Sigma^2} \left(\frac{L^2}{E^2} - \frac{\Xi^2}{\sin^2 \theta} (aE \sin^2 \theta - L)^2\right)}.
	\label{17}
	\end{equation}
	The azimuthal motion is unaffected by the photon’s  and is described as,
	\begin{equation}
	\dot{\phi} = \frac{\Xi}{\Sigma^2 \sin^2 \theta} \left(L - aE \sin^2 \theta\right).
	\label{18}
	\end{equation}
	The time motion for photons is described by,
	\begin{equation}
	\dot{t} = \frac{\Xi}{\Sigma^2} \left(E(r^2 + a^2) - aL\right).
	\label{19}
	\end{equation}
	These equations describe the motion of photons around the KNAdS BH and can be used to analyze the photon sphere and other photon dynamics in the presence of the BH's rotation and the quintessence parameter.
	
	To identify the critical and unstable circular orbits for photons around the KNAdS BH, we introduce the dimensionless parameters \(\eta = \frac{K}{E^2}\) and \(\xi = \frac{L}{E}\), where \(L\) is the angular momentum and \(E\) is the conserved energy of the photon. The conditions for critical and unstable circular orbits are derived from Eq. (\ref{16}) along with the condition (\ref{10}). The first condition that must be satisfied is that the radial potential vanishes, i.e., \(\dot{r}^2 = 0\), ensuring the photon remains at a fixed radial coordinate \(r_c\). The second condition is that the first derivative of the effective potential vanishes, \(\frac{\partial (\dot{r}^2)}{\partial r} = 0\), which determines the critical radius \(r_c\) where the orbit is circular. Finally, the sign of the second derivative of the effective potential determines the stability of the orbit.
	\begin{equation}
	\frac{\partial^2 (\dot{r}^2)}{\partial r^2} > 0 \quad \text{(stable circular orbit)},
	\label{20}
	\end{equation}
	\begin{equation}
	\frac{\partial^2 (\dot{r}^2)}{\partial r^2} < 0 \quad \text{(unstable circular orbit)}.
	\label{21}
	\end{equation}
	The radial potential for photons is given by,
	\begin{equation}
	R(r) = \frac{1}{\Sigma^2} \left[ \left( (r^2 + a^2)E - a L \right)^2 - \Delta_r \left( \eta + (L - aE)^2 \right) \right].
	\label{22}
	\end{equation}
	For circular orbits, the following conditions hold,
	\begin{equation} 
	R(r_c) = 0,~~ \frac{dR(r)}{dr}\bigg|_{r = r_c} = 0.
	\label{23}
	\end{equation}
	These conditions are used to solve for the critical radius \(r_c\) of the photon sphere and the associated values of \(\xi\) and \(\eta\). To find the equations for \(\xi\) and \(\eta\), we use the condition (\ref{23}), which ensures that the radial potential at the critical radius \(r_c\) vanishes, we get,
	\begin{equation}
	\frac{\partial}{\partial r} \left[ \left( (r^2 + a^2) - a \xi \right)^2 - \Delta_r \left( \eta + (\xi - a)^2 \right) \right] \Bigg|_{r = r_c} = 0.
	\label{24}
	\end{equation}
	Expanding and solving the above equation gives,
	\begin{equation}
	\xi = \frac{(r_c^2 + a^2) \Delta_r' - 4 r_c \Delta_r}{a \Delta_r' - 4 a r_c},
	\label{25}
	\end{equation}
	\begin{equation}
	\eta = \frac{r_c^2 \left[ 4 a^2 \Delta_r - (r_c^2 + a^2)^2 \Delta_r' \right]}{a^2 \Delta_r' - 4 a^2 r_c}.
	\label{26}
	\end{equation}
	where $\Delta_r' = \frac{d\Delta_r}{dr} \big|_{r = r_c}$.
	
	These are the required values of $\xi$ and $\eta$ at the circular orbit radius $r_c$. Using them and incorporating Eq. (\ref{17}), we can find the following inequality of photon region,
	\begin{equation}
	\begin{aligned}
	&a (r_c^2 - r^2)^2 (\Delta'_r)^3 - 8a (r_c^2 - r^2) r \Delta_r (\Delta'_r)^2 \\
	&+ 16a r^2 (\Delta_r\Delta'_r)^2- (r^2 + a^2)(\Delta'_r) + 4r\Delta_r > 0,
	\label{27}
	\end{aligned}
	\end{equation}
	where  $\Delta'_r$ is given by,
	\begin{equation}
	\Delta_r' = -2M + \frac{2a^2 r}{\ell^2} + 3\alpha\omega r^{-3\omega} - \alpha r^{-3\omega} + 2r + \frac{4r^3}{\ell^2}.
	\label{28}
	\end{equation}
	This inequality (\ref{27}) defines the condition for photon orbits in the KNAdS BH region. The  Fig. 2 illustrates the structural changes in the KNAdS BH with quintessence energy under varying parameters, demonstrating the ergosphere, static limit, photon region, and event horizons. Each row captures distinct regimes by altering the BHs spin parameter (\(a\)), mass (\(M\)), quintessence parameter (\(\alpha\)), and the equation of state parameter (\(\omega\)). The purple regions represent the ergosphere, where particles cannot remain stationary, while the red regions denote the static region. The white regions near the BH center correspond to the photon region, and the green/yellow contours indicate critical surfaces like the static limit and event horizons.
	
	In the first row, the static limit (outer green contour) is significantly influenced by the spin (\(a\)) and quintessence (\(\alpha\)). At lower spins, the static limit and ergosphere are nearly spherical, consistent with a lower deformation of spacetime. However, as \(a\) increases, the ergosphere elongates along the equatorial plane, and the static limit expands outward, as seen in the transition from left to right. The photon region (white) near the center becomes more asymmetric, reflecting the effects of spin and quintessence on null geodesics, consistent with KNAdS geometry in the literature.
	
	In the second row, the ergosphere exhibits more pronounced distortions, particularly at higher spins (\(a\)) and larger quintessence parameters (\(\alpha\)). The photon region shrinks or splits into multiple components, depending on the interplay of BH parameters. The static limit shifts outward, creating a larger region where particles cannot remain stationary, a typical feature in highly spinning KNAdS BH. By the third row, extreme parameter choices (\(a\), \(\alpha\), and \(\omega\)) lead to highly deformed ergospheres and disconnected photon regions. These configurations demonstrate the sensitivity of BH geometry to the quintessence field, showing stretched or compressed regions in accordance with modifications from the cosmological scalar field. The figure emphasizes how increasing spin and quintessence can radically alter the geometry of KNAdS BH, consistent with the predictions of modified spacetime metrics in the literature.
	\begin{figure*}[htbp]
		\centering
		\begin{minipage}{0.24\textwidth}
			\centering
			\includegraphics[width=\linewidth]{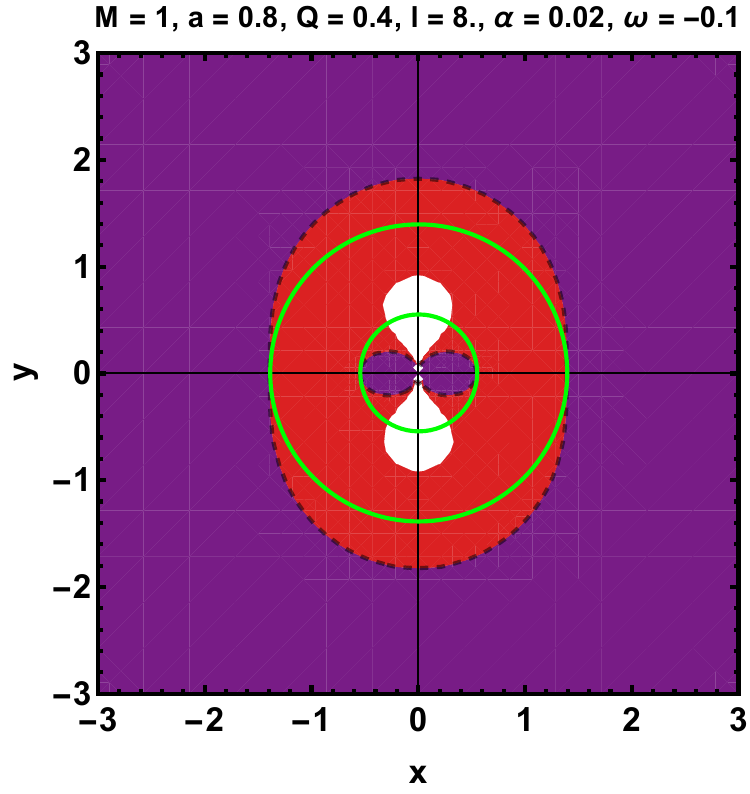}
		\end{minipage}%
		\begin{minipage}{0.24\textwidth}
			\centering
			\includegraphics[width=\linewidth]{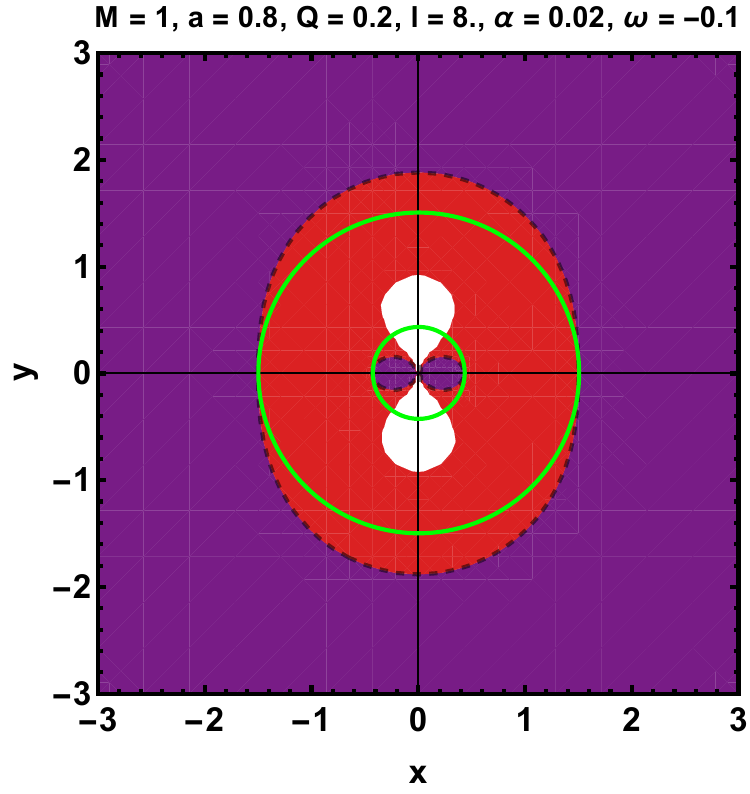}
		\end{minipage}%
		\begin{minipage}{0.24\textwidth}
			\centering
			\includegraphics[width=\linewidth]{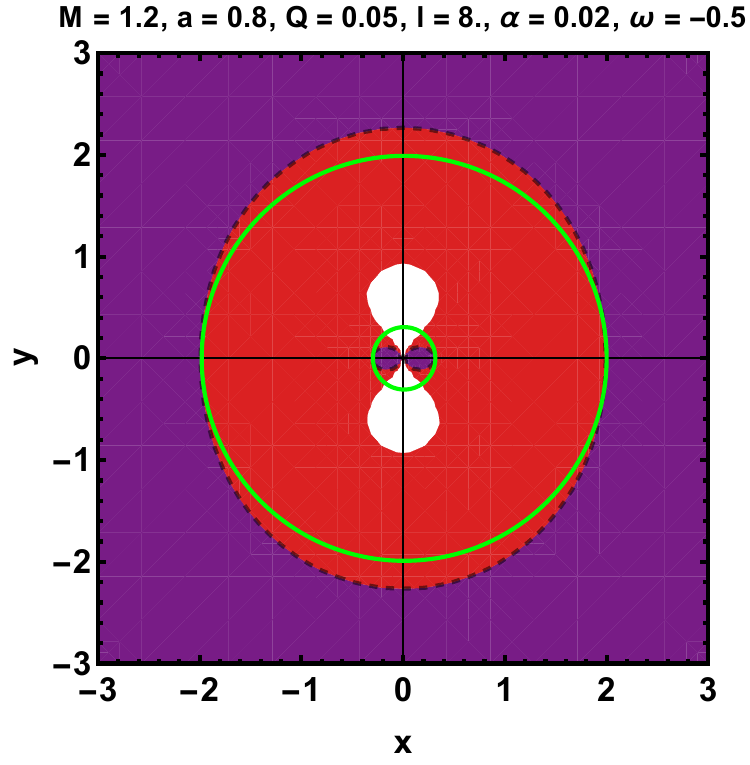}
		\end{minipage}%
		\begin{minipage}{0.24\textwidth}
			\centering
			\includegraphics[width=\linewidth]{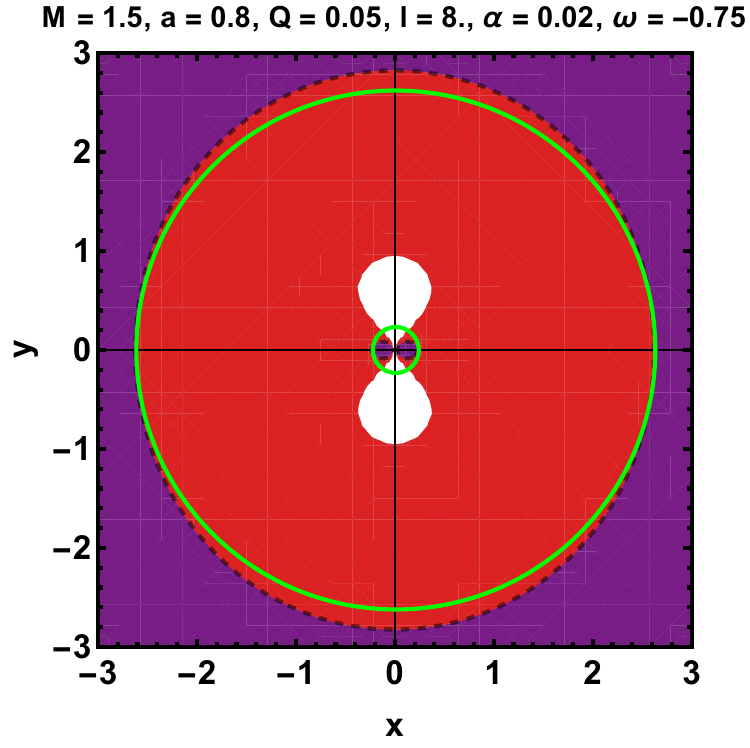}
		\end{minipage} \\
		
		\begin{minipage}{0.24\textwidth}
			\centering
			\includegraphics[width=\linewidth]{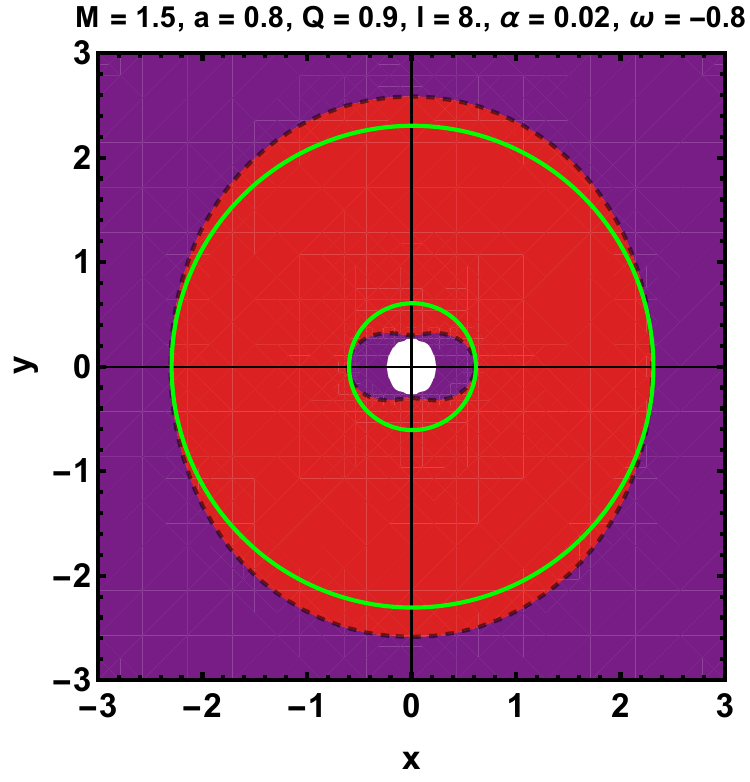}
		\end{minipage}%
		\begin{minipage}{0.24\textwidth}
			\centering
			\includegraphics[width=\linewidth]{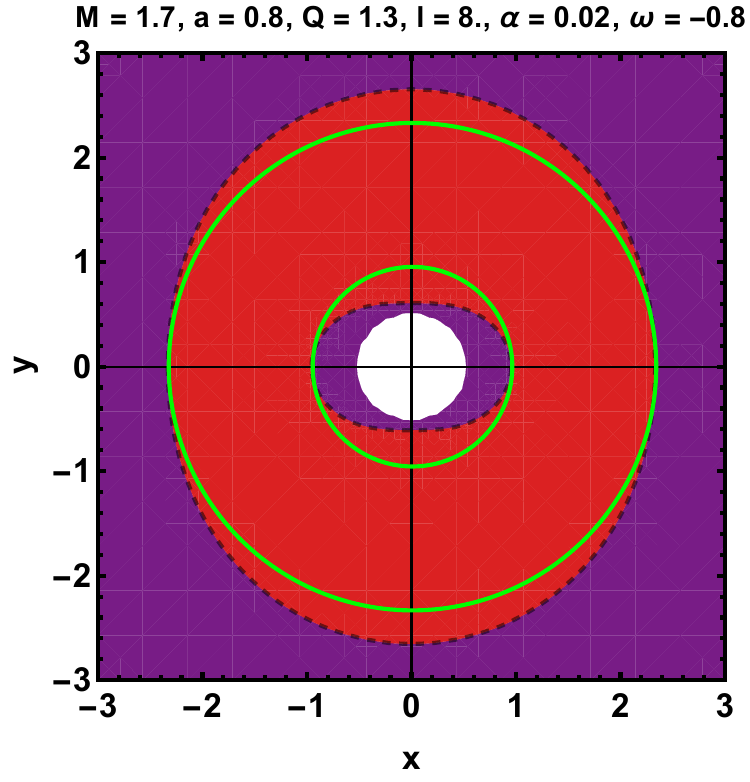}
		\end{minipage}%
		\begin{minipage}{0.24\textwidth}
			\centering
			\includegraphics[width=\linewidth]{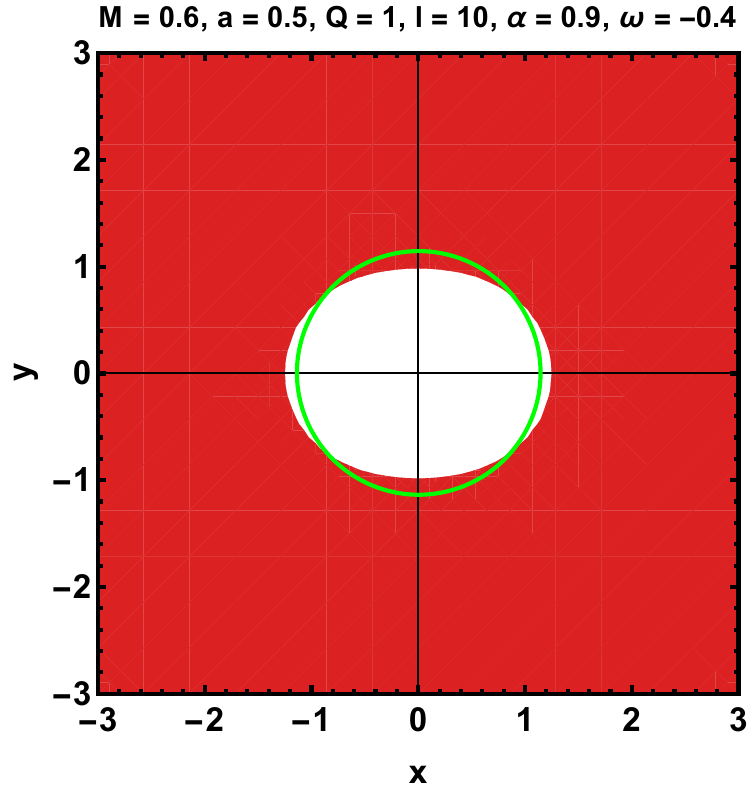}
		\end{minipage}%
		\begin{minipage}{0.24\textwidth}
			\centering
			\includegraphics[width=\linewidth]{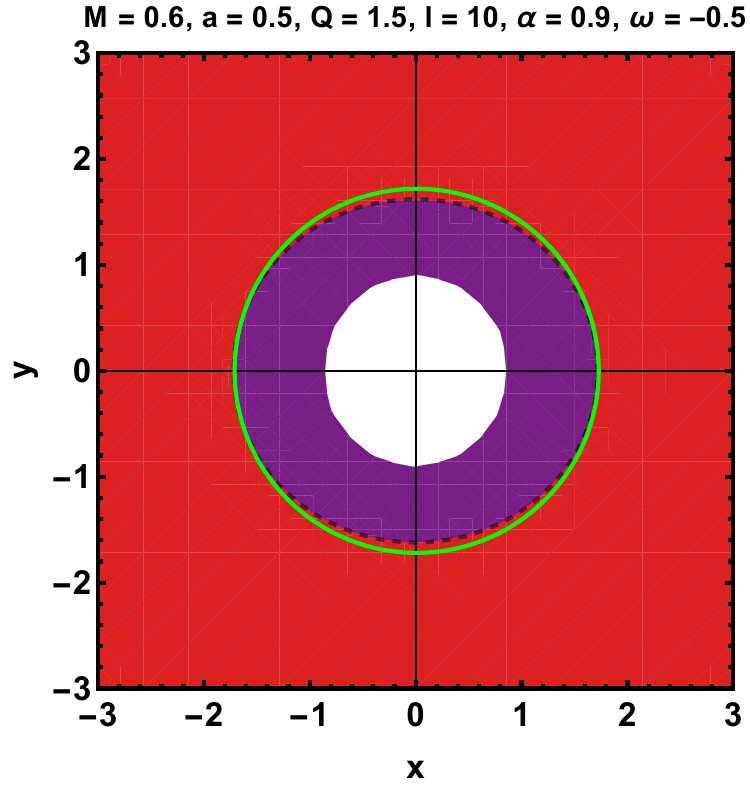}
		\end{minipage}
		\\
		
		\begin{minipage}{0.24\textwidth}
			\centering
			\includegraphics[width=\linewidth]{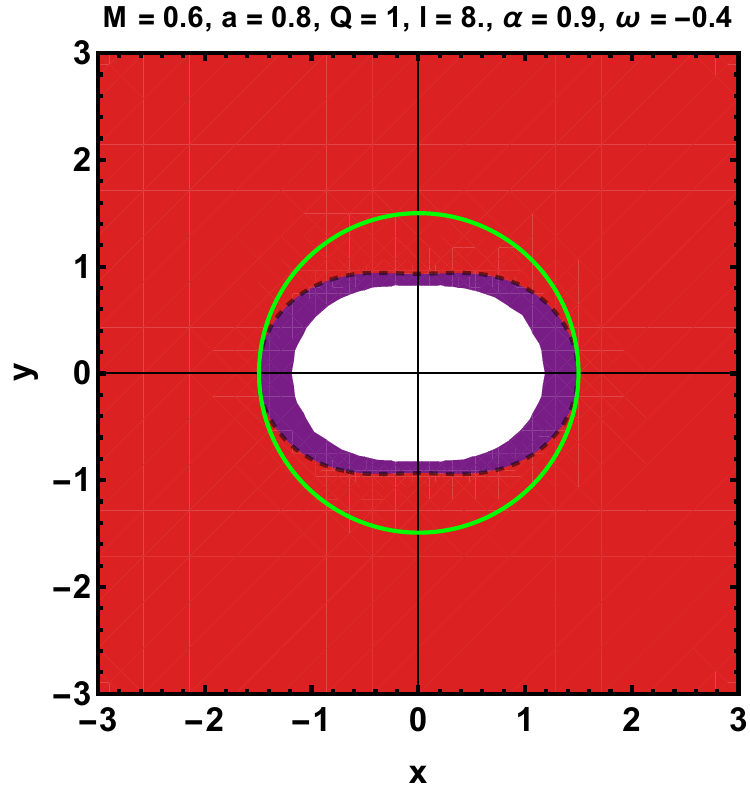}
		\end{minipage}%
		\begin{minipage}{0.24\textwidth}
			\centering
			\includegraphics[width=\linewidth]{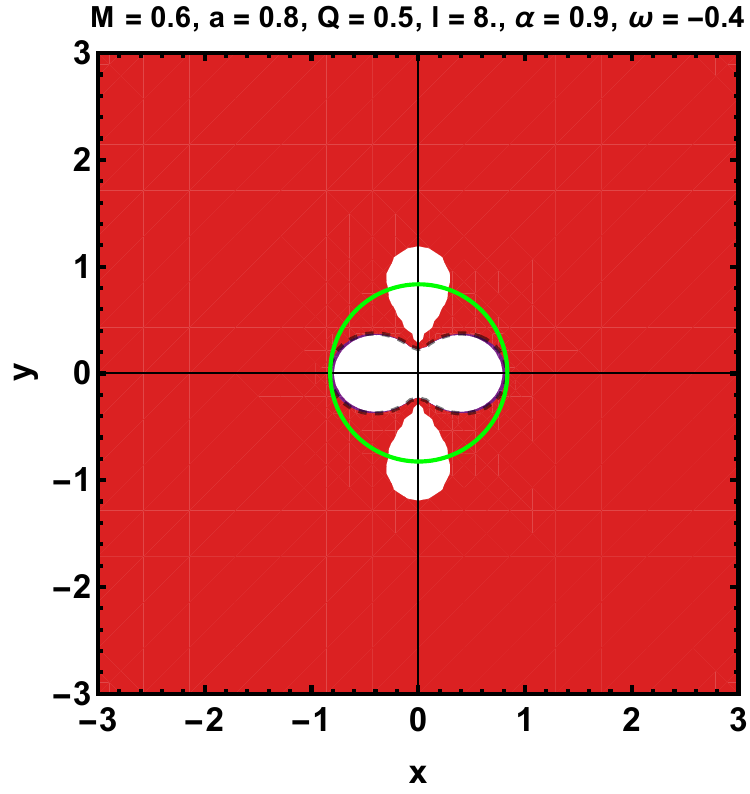}
		\end{minipage}%
		\begin{minipage}{0.24\textwidth}
			\centering
			\includegraphics[width=\linewidth]{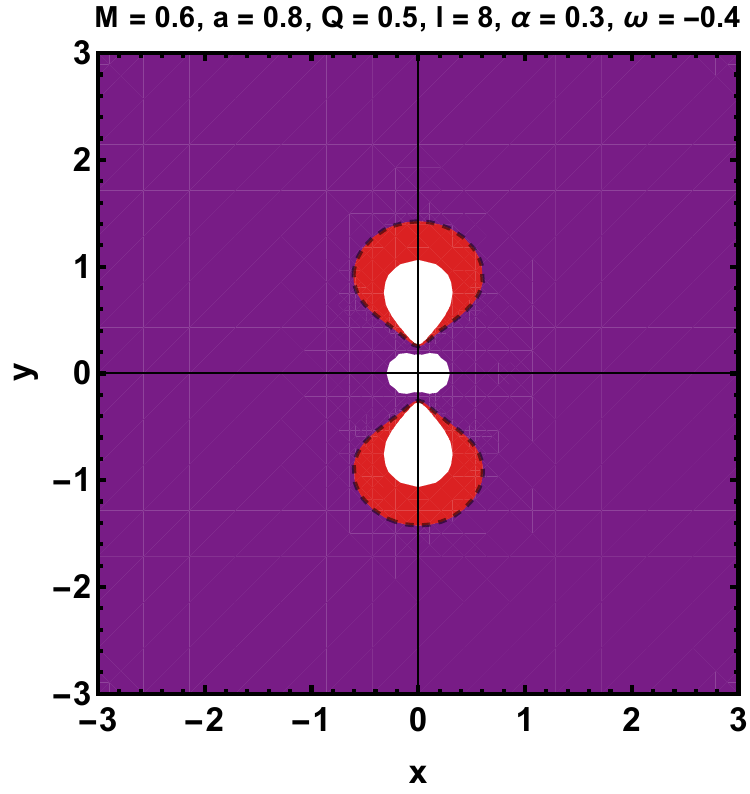}
		\end{minipage}%
		\begin{minipage}{0.24\textwidth}
			\centering
			\includegraphics[width=\linewidth]{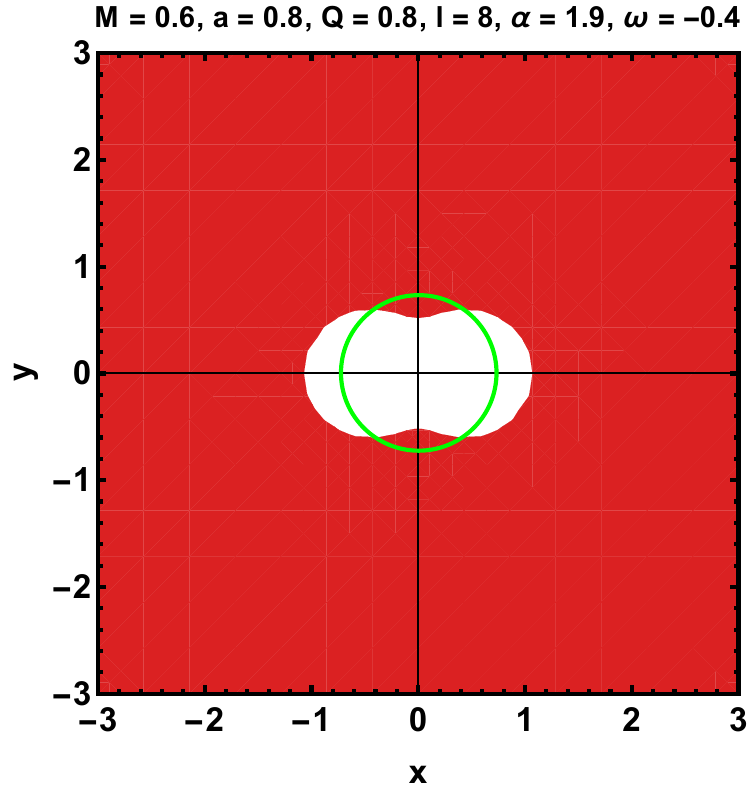}
		\end{minipage}
		\caption{Multiple contour plots of the \(g_{tt}\) component of the KNAdS metric in the presence of quintessence matter for varying quintessence state parameters \(\omega\) and quintessence strength \(\alpha\). Each subplot shows the \(g_{tt}\) contours in the \(x\)-\(y\) plane (\(x = r \cos\theta, y = r \sin\theta\)), with the static limit surface (\(g_{tt} = 0\)) as the dashed black contour and the horizons (roots of \(\Delta_r = 0\)) highlighted by green circles. The figures explores the effects of quintessence on the horizon structure, ergoregion, and static limit surface for different parameter combinations.} \label{fig:1}
	\end{figure*}
	\subsection{ Angular Velocity of a Particle in Ergoshpere}
	In this section, we aim to investigate the particle angular velocity (\(\Omega = \frac{d\phi}{dt}\)) and its limitations within the premises of the ergosphere under the condition \(ds^2 \geq 0\). Hence,  
	\begin{equation}
	g_{tt} \, dt^2 + g_{t\phi} \, dt \, d\phi + g_{\phi\phi} \, d\phi^2 \geq 0,
	\label{29}
	\end{equation}  
	and the angular velocity must satisfy the constraints \(\Omega_+ \leq \Omega \leq \Omega_-\) \cite{bardeen1973four}, where  
	\begin{equation}
	\Omega_\pm = \frac{-g_{t\phi} \pm \sqrt{g_{t\phi}^2 - g_{tt}g_{\phi\phi}}}{g_{\phi\phi}}.
	\label{30}
	\end{equation}  
	
	On the boundary of the ergosphere, \(g_{tt} = 0\) and \(\Omega_+ = 0\), while inside the ergosphere, \(g_{tt} < 0\) and \(\Omega_\pm > 0\), and every particle moves in the direction of the KNAdS BH rotation \cite{banados1998kerr}.  
	
	In the case of a rotating BH, substituting the values of \(g_{tt}\), \(g_{t\phi}\), and \(g_{\phi\phi}\) from Eq. (\ref{01}) into Eq. (\ref{29}), we obtain the angular velocity of a test particle as
	\begin{equation}
	\Omega_\pm = \frac{-\Delta_r a \sin^2\theta \pm \sqrt{\Delta_r \Delta_\theta} \sin\theta (r^2 - a^2)}{\Delta_\theta \sin^2\theta (r^2 - a^2)^2 - \Delta_r a^2 \sin^4\theta}.
	\label{31}
	\end{equation}
	As the particle approaches the event horizon (\(r \to r_+\)), we find that,
	\begin{equation}
	\lim_{r \to r_{+}} \Omega_{+} = \lim_{r \to r_{+}} \Omega_{-} = \omega_{bh} = \frac{\sqrt{(r - r_+)} \sin\theta (r^2 - a^2)}{\Delta_\theta \sin^2\theta (r^2 - a^2)^2},
	\label{32}
	\end{equation}
	where \(\omega_{bh}\) denotes the angular velocity of the KNAdS BHs rotation.
	
	The specific angular momentum \(L\) and specific energy \(E\) for circular orbits are derived from the effective potential conditions \(V_{\text{eff}}(r) = 0\) and \(\frac{dV_{\text{eff}}(r)}{dr} = 0\). 
	To find the effective potential for geodesic motion in this spacetime, solving for \(\dot{r}^2\) gives the radial motion equation of the form,  
	\begin{equation}
	\dot{r}^2 = E^2 - V_{\text{eff}}(r, \theta),
	\label{33}
	\end{equation}
	where \(V_{\text{eff}}\) is the effective potential.
	Incorporating Eq. (\ref{16}) we have,
	\begin{equation} V_{\text{eff}} = \frac{L^2}{E^2} + \frac{\Delta_r\left(E(r^2 + a^2) - aL\right)^2}{(r^2 + a^2 \cos^2 \theta)}. 
	\end{equation}
	Using the  conditions for effective potential, the explicit expressions for \(E\) and \(L\) are:
	\begin{equation}
	E = \sqrt{\frac{\Delta_r \mathcal{H}^2}{(r^2 + a^2)^2 - 4a^2 r^2}},
	\label{34}
	\end{equation}
	and
	\begin{equation}
	L = \frac{\sqrt{\Delta_r} \left[ a (r^2 + a^2)\mathcal{H} + r^2 \sqrt{(r^2 + a^2)^2 - 4a^2 r^2} \right]}{(r^2 - 2a^2) \sqrt{(r^2 + a^2)^2 - 4a^2 r^2}},
	\label{35}
	\end{equation}
	where
	\begin{equation}
	\mathcal{H} = \sqrt{r^2 + a^2 - 2Mr}.
	\label{36}
	\end{equation}
	Now, the angular velocity \(\Omega\) of the particle orbiting a rotating BH is defined as the ratio of angular momentum \(L\) to energy \(E\). Thus, the expression for \(\Omega\) becomes:
	
	\begin{equation}
	\Omega = \frac{\mathcal{H} a (r^2 + a^2) + r^2 \sqrt{(r^2 + a^2)^2 - 4a^2 r^2}}{(r^2 - 2a^2) \Delta_r}.
	\label{37}
	\end{equation}
	
	In the literature, the expressions for the angular velocity \(\Omega\), energy \(E\), and angular momentum \(L\) of a test particle in a Kerr BH's ergosphere are well-established. These equations demonstrate the influence of the black hole's rotation on particle motion, specifically within the ergosphere, where the particles are forced to co-rotate with the BH.
	
	For instance, the study by Bardeen, Carter, and Hawking \cite{bardeen1973four} provides a comprehensive analysis of the motion of particles around rotating BHs. The critical observation that particles in the ergosphere are forced to move in the direction of the BHs rotation aligns with the above expressions. The formula for \(\Omega\) derived here is consistent with the earlier results for a Kerr BH, where \(\Omega\) is bounded within the ergosphere, and particles are constrained to co-rotate with the BH as they approach the event horizon \cite{chandrasekhar1983mathematical}.
	
	The values of \(E\) and \(L\) depend on the particle's radius and the BHs spin parameter, confirming that energy and angular momentum are not conserved in the same way in the ergosphere as they would be in a static spacetime, due to the frame-dragging effect. Thus, the inclusion of \(\alpha\), \(\omega_{bh}\), and \(L\) is crucial for a comprehensive understanding of particle dynamics in the ergosphere of a rotating black hole, and the derived expressions are consistent with the well-established framework in GR.
	\begin{figure*}[t]
		\centering
		\begin{minipage}[t]{0.4\textwidth}
			\centering
			\includegraphics[width=1\linewidth]{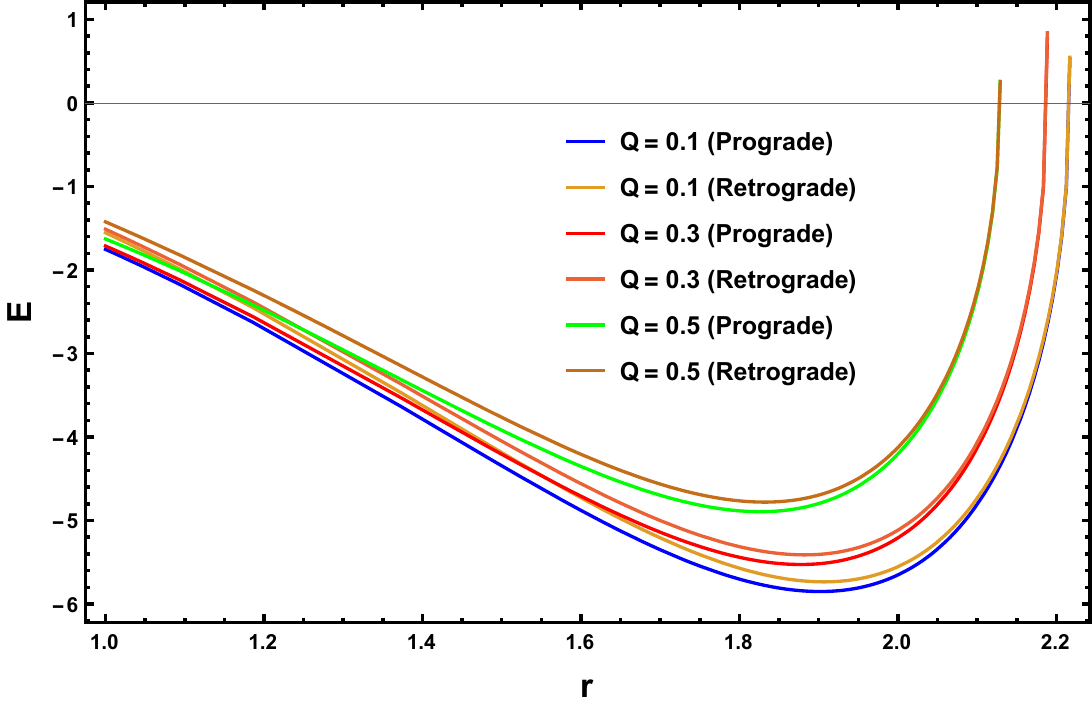}
			Fig. 3(a)
		\end{minipage}%
		\hspace{0.05\textwidth}
		\begin{minipage}[t]{0.4\textwidth}
			\centering
			\includegraphics[width=1\linewidth]{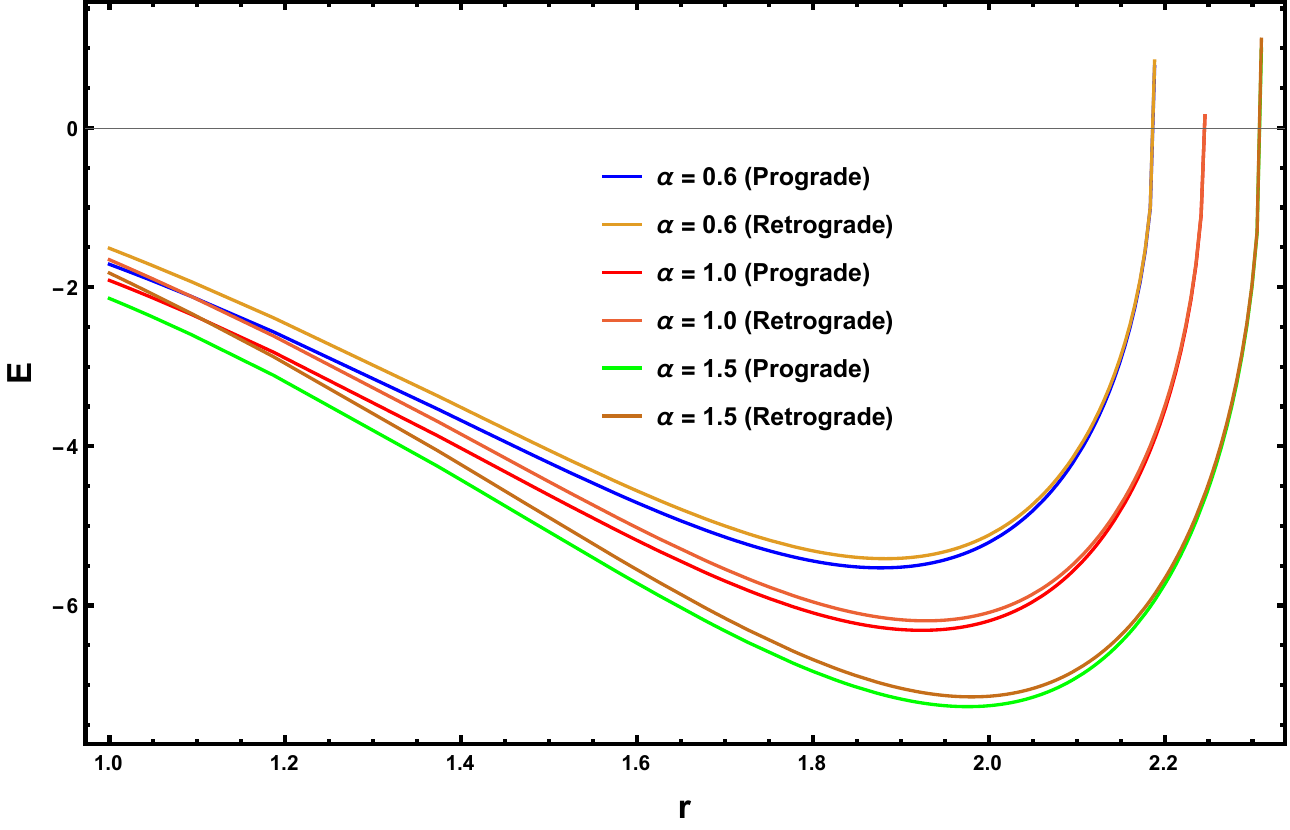}
			Fig. 3(b)
		\end{minipage}
		
		\vspace{0.5cm}
		\begin{minipage}[t]{0.4\textwidth}
			\centering
			\includegraphics[width=1\linewidth]{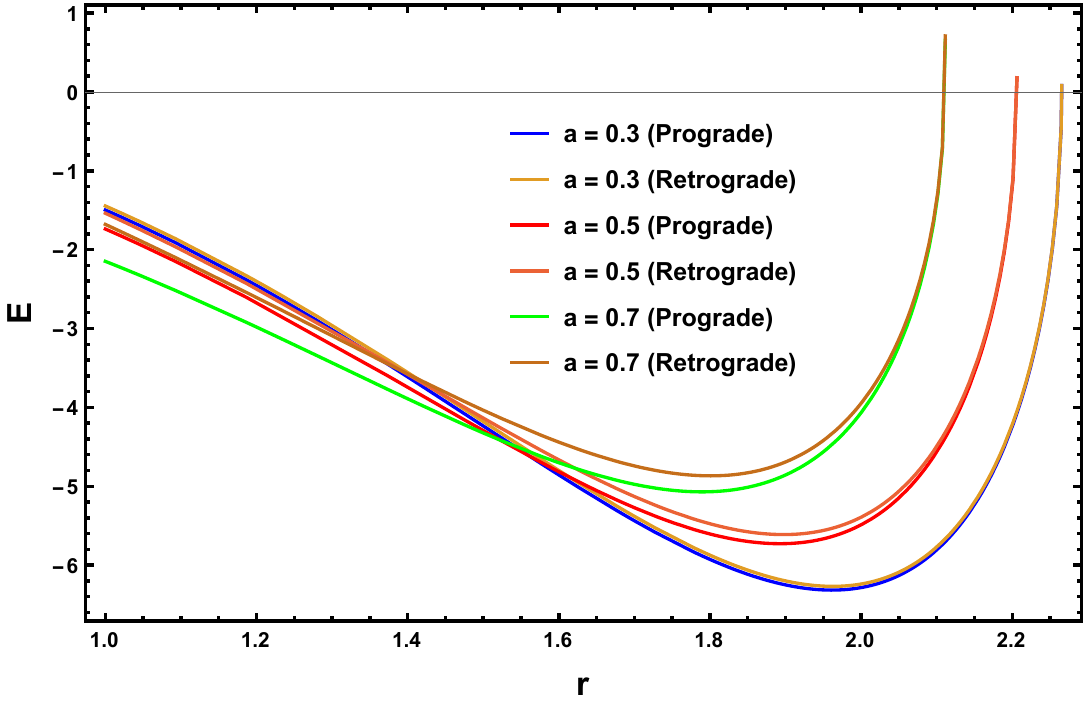}
			Fig. 3(c)
		\end{minipage}%
		\hspace{0.05\textwidth}
		\begin{minipage}[t]{0.4\textwidth}
			\centering
			\includegraphics[width=1\linewidth]{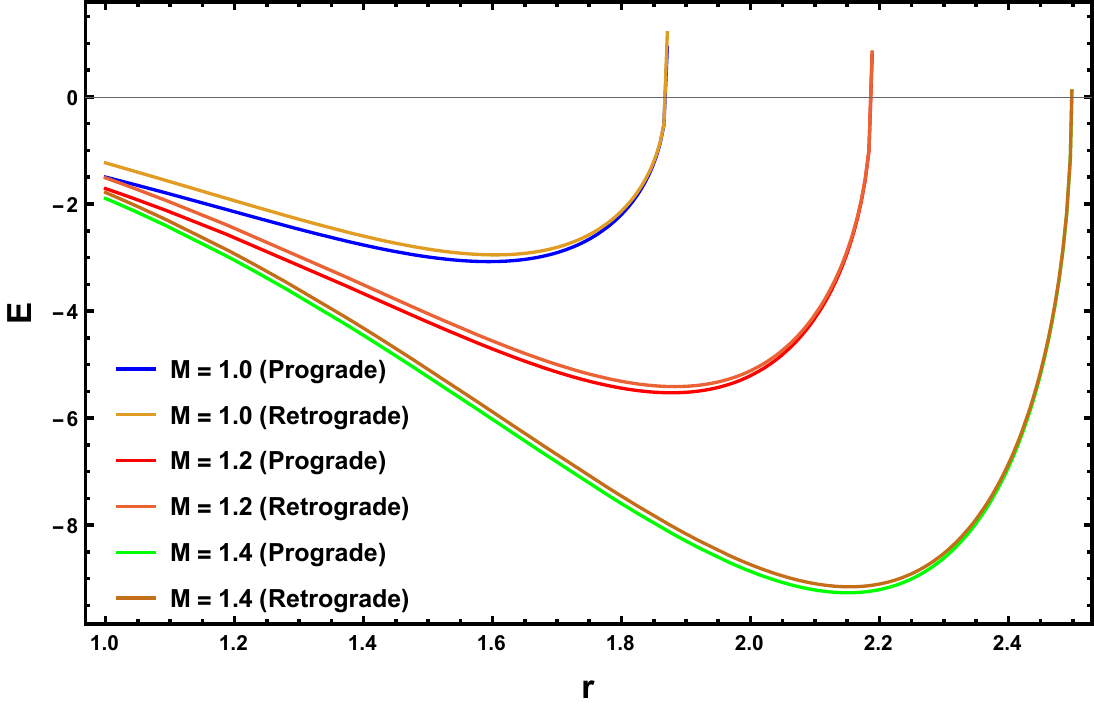}
			Fig. 3(d)
		\end{minipage}
		
		\captionsetup{justification=raggedright,singlelinecheck=false}
		\caption{Energy profiles \( E \) for test particles around a KNAdS BH, illustrating the effects of spin parameter \( a \), BH mass \( M \), and radial distance \( r \). Figures 3(a)–3(c) show the specific energy \( E \) for prograde (solid lines) and retrograde (dashed lines) orbits with varying spin parameters (\( a = 0.1\) to \( 0.9 \)), highlighting the increasing asymmetry and decreasing energy for prograde motion as \( a \) increases. Figure 3(d) demonstrates the influence of black hole mass (\( M = 1.0, 1.2, 1.4 \)) on \( E \), revealing that higher \( M \) lowers energy requirements and shifts the minima of \( E \) outward. These plots showcase the interplay between BH spin, mass, and spacetime geometry in governing orbital dynamics.}
		\label{fig:energy_profiles}
	\end{figure*}
	\begin{figure*}[t]
		\centering
		\begin{minipage}[t]{0.4\textwidth}
			\centering
			\includegraphics[width=1\linewidth]{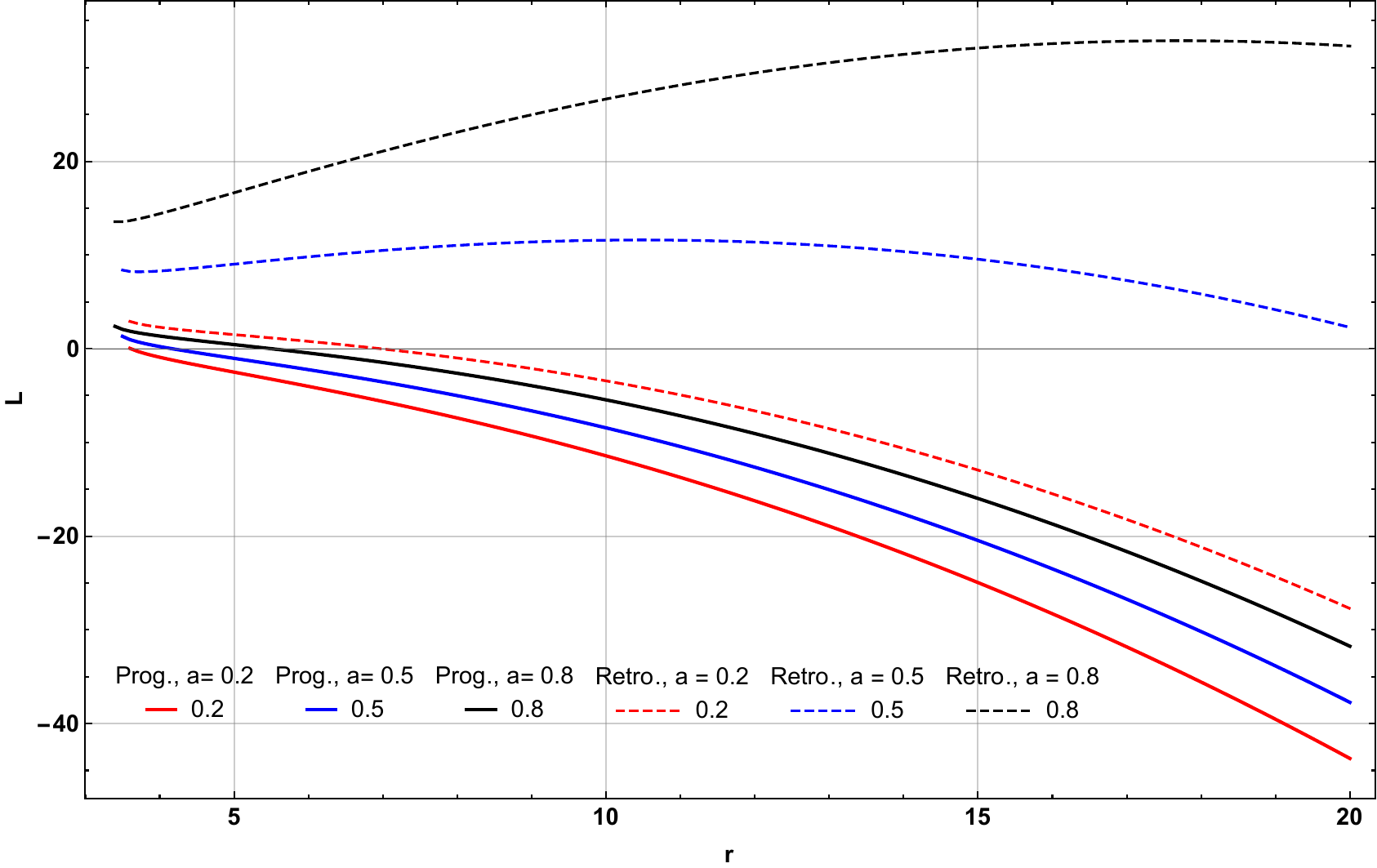}
			Fig. 4(a)
		\end{minipage}%
		\hspace{0.05\textwidth}
		\begin{minipage}[t]{0.4\textwidth}
			\centering
			\includegraphics[width=1\linewidth]{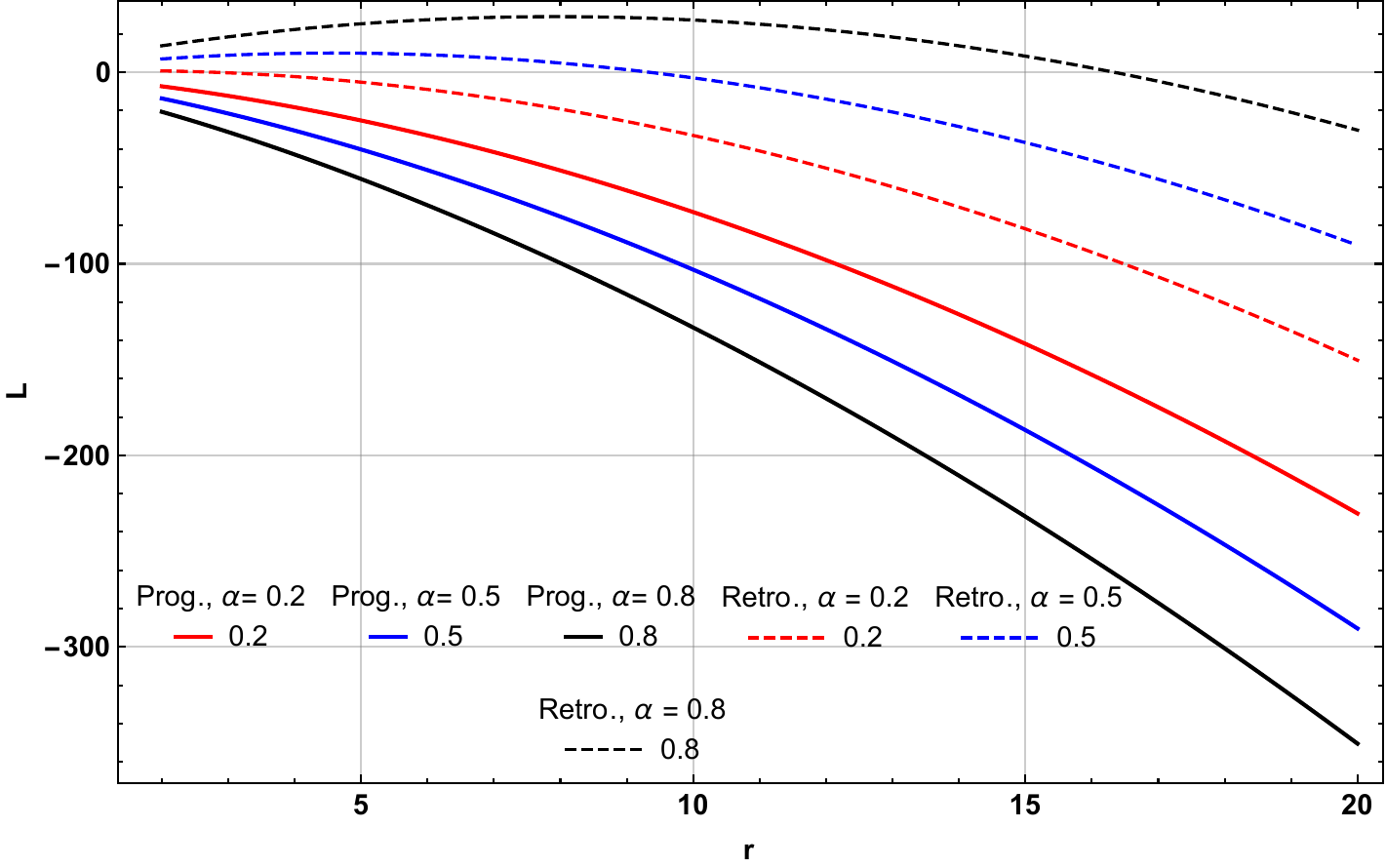}
			Fig. 4(b)
		\end{minipage}
		
		\vspace{0.5cm}
		\begin{minipage}[t]{0.4\textwidth}
			\centering
			\includegraphics[width=1\linewidth]{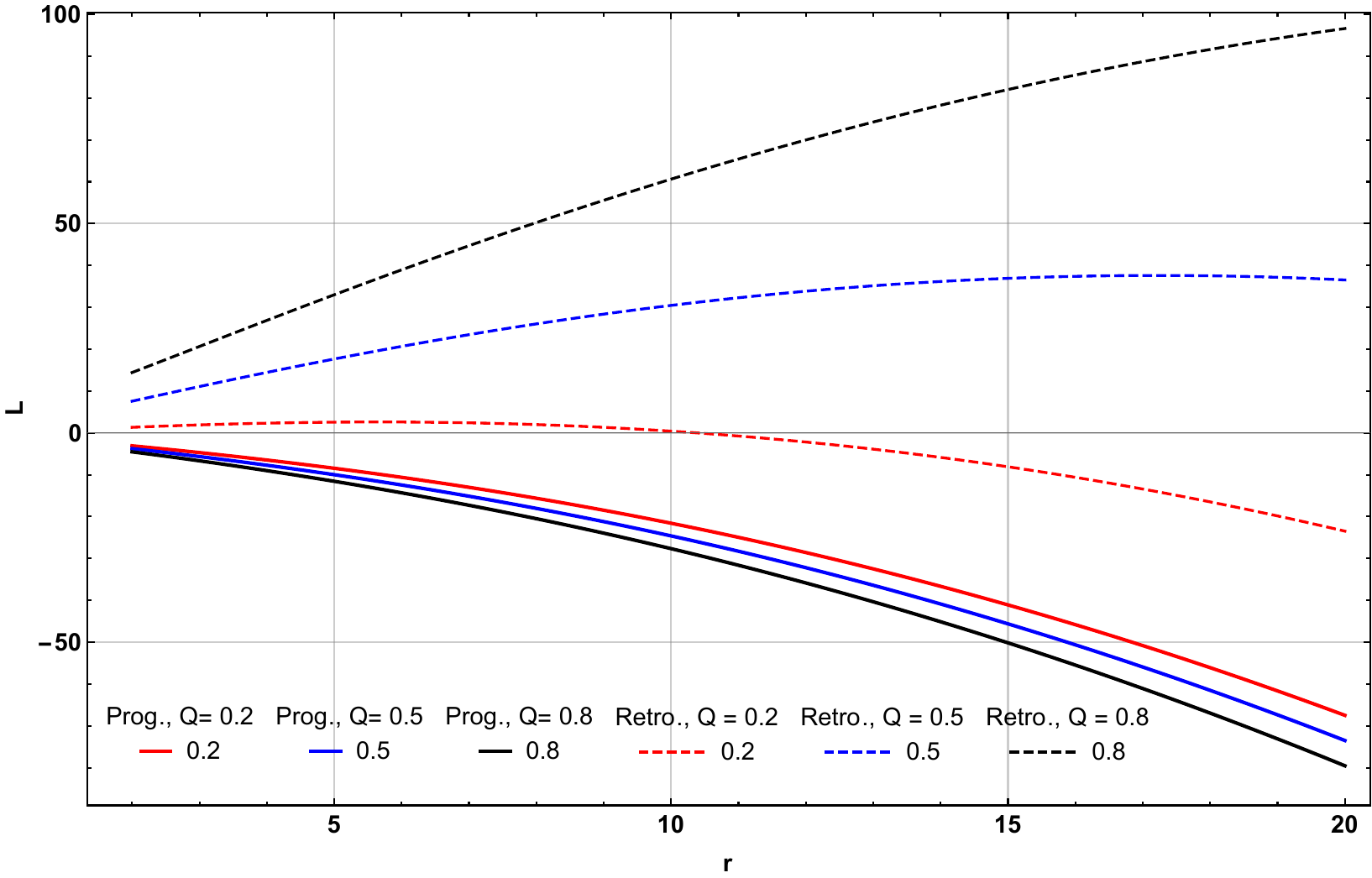}
			Fig. 4(c)
		\end{minipage}%
		\hspace{0.05\textwidth}
		\begin{minipage}[t]{0.4\textwidth}
			\centering
			\includegraphics[width=1\linewidth]{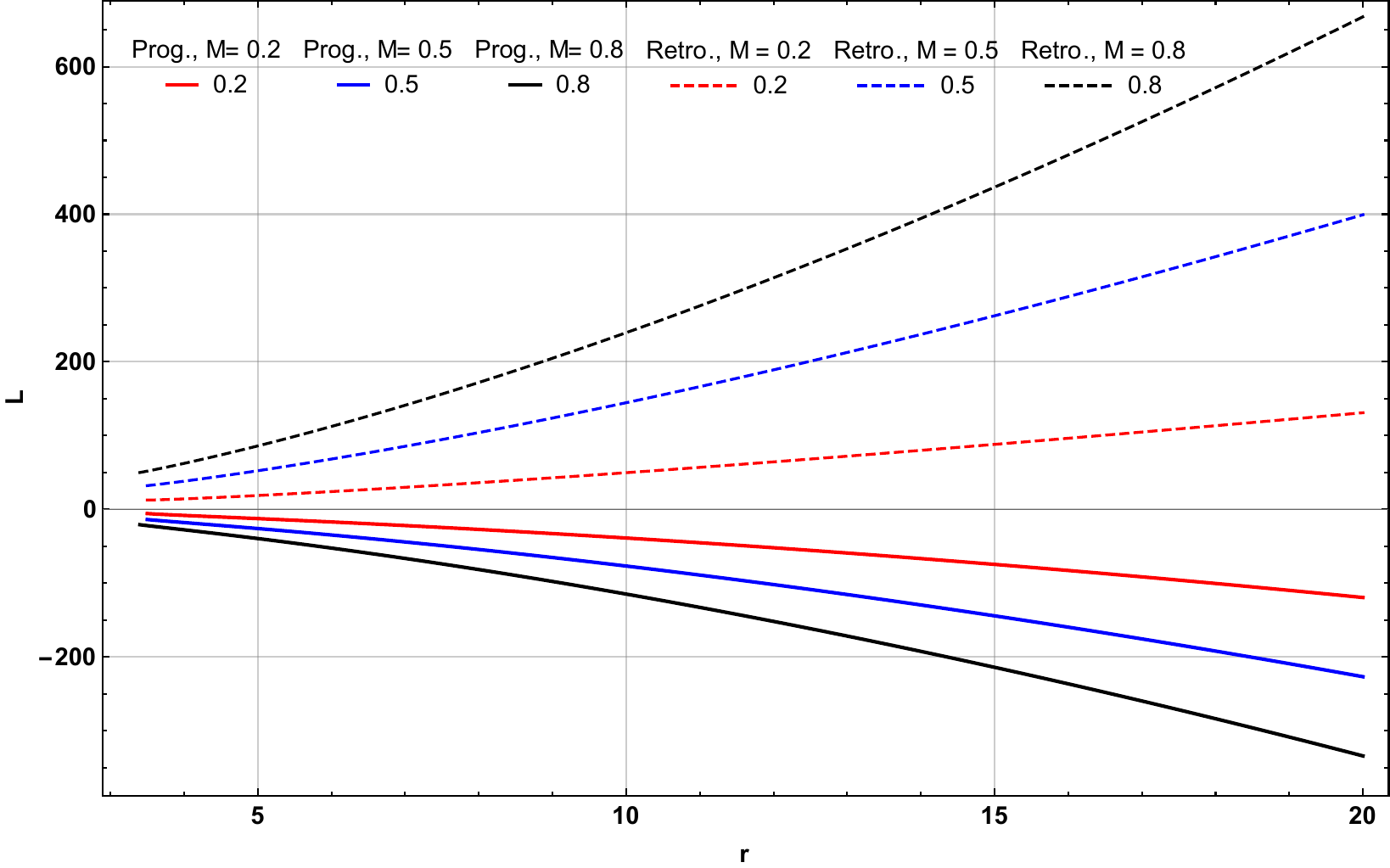}
			Fig. 4(d)
		\end{minipage}
		
		\captionsetup{justification=raggedright,singlelinecheck=false}
		\caption{Angular momentum profiles for prograde and retrograde orbits around a KNAdS BH with quintessence matter for varying spin parameter \( a \), black hole charge \( Q \), and quintessence parameter \( \alpha \). Figures 4(a)–4(d) illustrate the interplay between specific angular momentum \( L \) and energy \( E \), demonstrating the distinct behaviors of prograde (solid lines) and retrograde (dashed lines) orbits under different parameter settings.}
		\label{fig:energy_profiles}
	\end{figure*}
	
	Prograde motion refers to the trajectory of a test particle orbiting in the same direction as the spin of a KNAdS BH. This motion is strongly influenced by the frame-dragging effect, a relativistic phenomenon where the rotating BH drags the surrounding spacetime in the direction of its rotation. Due to this effect, prograde orbits require less specific energy \( E \), as the BH’s spin contributes positively to the particle’s angular momentum. The energy profiles in Fig. 3(a)–(d) reveal that as the spin parameter \( a \) increases, the energy \( E \) for prograde orbits decreases, allowing the particle to orbit closer to the event horizon. This behavior demonstrates the energetic advantage provided by the BH’s rotation. The reduced energy requirements for prograde orbits highlight the interplay between rotational dynamics and spacetime curvature, a hallmark feature of GR in strong gravitational fields.
	
	In contrast, retrograde motion describes a test particle orbiting against the direction of the BH’s spin. Here, the frame-dragging effect works against the particle’s motion, requiring higher specific energy $E$ to maintain orbit. The energy profiles in the figures clearly show that retrograde orbits are energetically less favorable compared to prograde ones, with $E$ increasing as $a$ grows. Moreover, retrograde orbits are restricted to regions further from the black hole, as the counter-rotating spacetime exerts a repelling effect on the particle’s motion. This energy asymmetry between prograde and retrograde orbits underscores the profound influence of rotational effects on orbital dynamics near BHs. The analysis illustrates how BH spin and its interaction with surrounding spacetime significantly alter the energy landscape for test particles, providing critical insights into the complex behavior of matter in extreme astrophysical environments.
	
	The plots of angular momentum $L$ in Fig. 4 illustrate the dynamics of particle motion around a KNAdS BH influenced by quintessence matter, characterized by the quintessence parameter and the spin. The prograde orbits (solid lines) exhibit lower specific angular momentum $L$ for a given energy $E$, compared to retrograde orbits (dashed lines). This is because the prograde motion aligns with the BH's rotational direction, benefiting from the frame-dragging effect, which reduces the effective angular momentum required for stable orbits. Conversely, retrograde motion opposes the rotation, requiring higher $L$ for the same $E$, reflecting the additional resistance induced by the counter-rotating spacetime geometry.
	
	In the presence of quintessence, the spacetime is further modified by an additional repulsive component depending on \( \alpha \). As \( \alpha \) increases, the effective potential is altered, influencing the stable orbit radii. The energy-angular momentum curves shift accordingly, indicating how quintessence weakens the gravitational pull, especially at large distances. These insights are crucial for understanding accretion disk dynamics, test particle motion, and astrophysical phenomena in the vicinity of rotating BHs embedded in a cosmological setting, including the interaction of dark energy with BH spacetimes.
	\subsection{ Effective Force and Lyapunov Exponents}
	The motion of information, such as whether it is directed away from or attracted toward a BH, can be described by an effective force. This force is related to the gradient of the effective potential acting on a photon, which can be expressed as:
	\begin{equation}
	F = \frac{-1}{2} \left( \frac{d V_{\text{eff}}}{d r} \right).
	\label{49}
	\end{equation}
	This equation reflects how the effective potential influences the dynamics of light or other test particles in curved spacetime. In the context of GR and BH physics, Lyapunov exponents are pivotal for analyzing the stability of geodesics and the chaotic behavior near BHs. Lyapunov exponents quantify the rate at which nearby trajectories—such as null or timelike geodesics—diverge or converge in the highly curved spacetime around a BH.
	
	The Lyapunov exponent $\lambda$ for a geodesic can be understood in terms of the growth rate of the separation $\|\xi^\mu\|$ between two nearby geodesics. In BH spacetimes, this exponent is associated with the instability of circular orbits. A general expression for the Lyapunov exponent \cite{Giri2022,Lei2022} is given by
	\begin{equation}
	\lambda = \sqrt{-\frac{V''_{\text{eff}}}{2}},
	\label{50}
	\end{equation}
	where $V_{\text{eff}}$ is the effective potential governing the motion of the particle or photon, and  $V''_{\text{eff}}$ is the second derivative of the effective potential with respect to the radial coordinate $r$. The negative sign ensures that $\lambda$ is real for unstable orbits, indicating that the geodesic motion exhibits sensitive dependence on initial conditions, a hallmark of chaotic behavior. The first derivative of $V_{\text{eff}}(r)$, is given as
	\begin{equation}
	\begin{aligned}
	V^{'}_{\text{eff}}(r) &= (r^2 + a^2 \cos^2 \theta) \big[\Delta_r' \left(E(r^2 + a^2) - aL\right)^2\\
	&+2\Delta_r\left(E(r^2 + a^2) - aL\right) (2Er)\big] - 2r \Delta_r\\
	&\times\left(E(r^2 + a^2) - aL\right)^2
	\div(r^2 + a^2 \cos^2 \theta)^2,
	\end{aligned}
	\label{52}
	\end{equation}
	where ``prime"  denotes the derivative withe respect to $r$. Finally, the second derivative, \( V^{''}_{\text{eff}}(r) \), is given by,
	\begin{equation}
	\begin{aligned}
	V^{''}_{\text{eff}}(r) &= \Delta_r'' \left(E(r^2 + a^2) - aL\right)^2 + 2 \Delta_r' \left(E(r^2 + a^2) - aL\right)  \\
	&\times 2Er+ 2 \Delta_r' \left(E(r^2 + a^2) - aL\right)(2Er) + 8 \Delta_r  (Er)^2 \\
	&+ 2 \Delta_r \left(E(r^2 + a^2) - aL\right) \cdot 2E-2\big[\Delta_r' (E(r^2 + a^2)\\
	&-aL^2 )\div( r + \Delta_r (E(r^2 + a^2) - aL)^2 \cdot \frac{2}{r^2})\big],
	\end{aligned}
	\end{equation}
	where
	\begin{equation}
	\Delta_r'' = \frac{2a^2}{\ell^2} + \left(3\alpha\omega - 9\alpha\omega^2\right) r^{-3\omega - 1} + 2 + \frac{12r^2}{\ell^2}.
	\end{equation}
	This is central to the calculation of the Lyapunov exponent for the system. These expressions enable a detailed analysis of the behavior of geodesics near the  KNAdS BH and the potential for chaotic dynamics in such a spacetime.
	
	The analysis of the effective force (\(F\)) for the KNAdS BH with quintessence energy, as shown in Fig. 5(a) and (b), reveals the relationship between gravitational attraction, centrifugal forces, and the repulsive effects of quintessence. For increasing energy (\(E\)), the effective force weakens near smaller radii, indicating that more energetic particles can counteract the black hole’s gravitational pull and remain in less curved regions of spacetime. Similarly, as angular momentum (\(L\)) increases, the centrifugal force grows, reducing the steepness of the effective force and allowing particles to maintain stable orbits farther from the black hole. Quintessence energy, with its repulsive influence, modifies the effective potential, shifting these effects outward. The AdS boundary, on the other hand, acts as a confining mechanism, ensuring that particles do not escape to infinity and remain in bounded regions of spacetime.
	
	The Lyapunov exponent (\(\lambda\)), which quantifies orbital stability, highlights the chaotic behavior near the black hole, as illustrated in Fig. 6(a) and (b). For higher energy, the region of instability (positive \(\lambda\)) broadens and shifts outward, suggesting that energetic particles can explore chaotic orbits farther from the black hole. Increasing angular momentum  stabilizes orbits near the black hole by introducing a stronger centrifugal barrier, reducing chaos at smaller radii. However, higher \(L\) extends the chaotic region outward, indicating that angular momentum allows particles to enter instability zones farther from the black hole. Quintessence energy further enhances this outward shift due to its repulsive nature, while the AdS boundary confines the chaotic regions within finite spatial domains, amplifying instability in certain zones.
	\begin{figure*}[t]
		\centering
		\begin{minipage}[t]{0.38\textwidth}
			\centering
			\includegraphics[width=\linewidth]{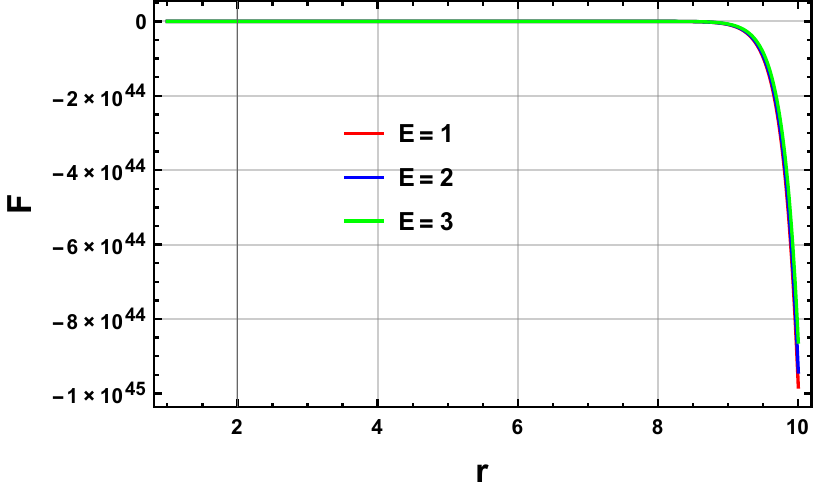}
			Fig. 5(a)
		\end{minipage}%
		\hspace{1.8cm} 
		\begin{minipage}[t]{0.38\textwidth}
			\centering
			\includegraphics[width=\linewidth]{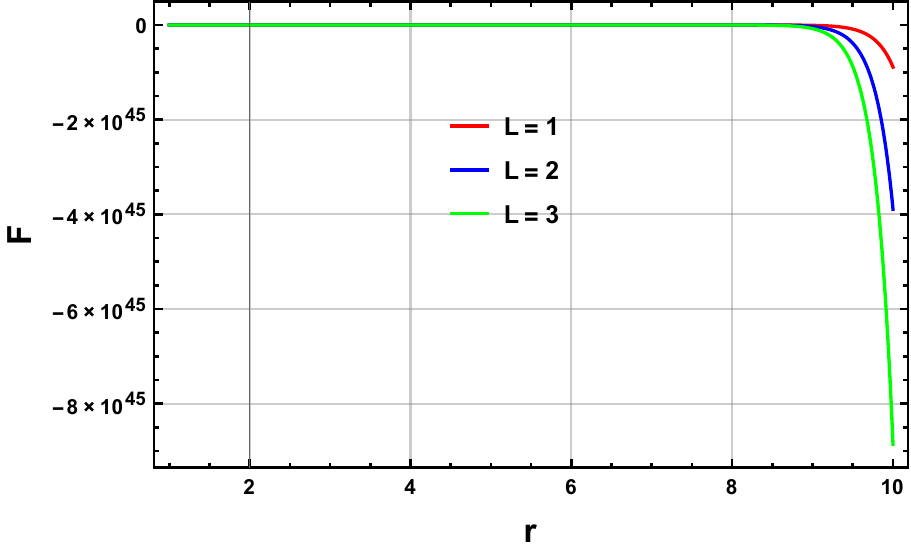}
			Fig. 5(b)
		\end{minipage}
		\captionsetup{justification=raggedright,singlelinecheck=false}
		\caption{Effective force (\(F\)) versus radial coordinate (\(r\)) for different energy values (\(E\)) in (a) and angular momentum values (\(L\)) in (b). Higher \(E\) and \(L\) reduce the steepness of the force near small radii.}
	\end{figure*}
	\begin{figure*}[t]
		\centering
		\begin{minipage}[t]{0.38\textwidth}
			\centering
			\includegraphics[width=\linewidth]{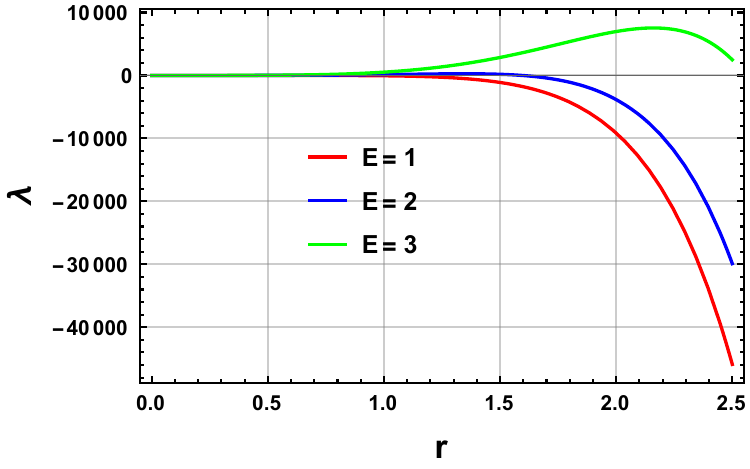}
			Fig. 6(a)
		\end{minipage}%
		\hspace{1.8cm} 
		\begin{minipage}[t]{0.38\textwidth}
			\centering
			\includegraphics[width=\linewidth]{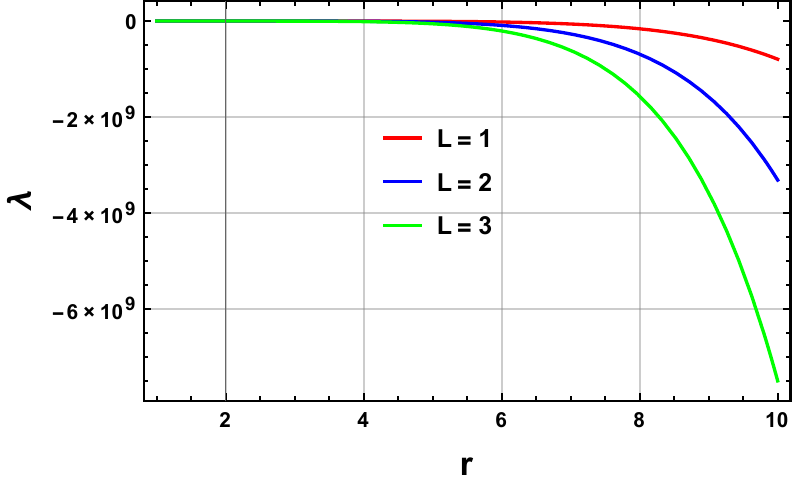}
			Fig. 6(b)
		\end{minipage}
		\captionsetup{justification=raggedright,singlelinecheck=false}
		\caption{Lyapunov exponents (\(\lambda\)) versus radial coordinate (\(r\)) for different energy values (\(E\)) in (a) and angular momentum values (\(L\)) in (b). Positive peaks indicate chaotic regions, with higher \(E\) and \(L\) shifting the instability regions outward}
	\end{figure*}
	
	\section{The Penrose Process}
	
	In this section, we discuss the Penrose mechanism and energy extraction from the ergoregion of a rotating black hole influenced by a quintessential energy field. As a cornerstone problem in GR, this mechanism facilitates the extraction of rotational energy from BHs, governed by the fundamental principles of energy and angular momentum conservation. It is considered one of the most efficient processes for energy extraction, surpassing even the efficiency of nuclear reactions. This remarkable process not only offers insights into high-energy astrophysical phenomena but also provides a deeper understanding of the interaction between BHs and surrounding fields.  
	
	The concept of energy extraction via the Penrose mechanism was first proposed by Roger Penrose in 1969 \cite{Penrose1969}, and subsequent studies have explored its implications in diverse astrophysical settings \cite{Carter1968, Christodoulou1971}. When a rotating BH is surrounded by a quintessential energy field, the dynamics of the process become even more intricate, as discussed in  \cite{Chandrasekhar1983}.  
	
	The efficacy of the Penrose mechanism hinges on two critical criteria: the absorption of angular momentum and the existence of negative energy states within the ergoregion. This investigation aims to elucidate the influence of the deviation parameter $\alpha$, on the emergence of negative energy states $E$, and the overall efficiency of the mechanism. Furthermore, the study explores the role of the rotational parameter $a$, analyzing its impact on the mechanism's efficiency and the behavior of negative energy states associated with the BH.
	
	\subsection{Negative Energy States}
	Negative energy states may arise due to counter-rotating orbits (as seen in the Kerr black hole spacetime) or due to electromagnetic interactions (as in the case of the Reissner-Nordström BH). These states are of significant interest, particularly for understanding energy extraction mechanisms, such as the Penrose process, and for probing the limits of particle energy at specific spacetime locations. The energy and angular momentum of a particle play a crucial role in determining these states. The radial equation governing the motion of a particle in this spacetime can be rewritten as
	\begin{equation}
	\Delta_r (r^2 + a^2)^2 E^2 - 2aE L \Delta_r (r^2 + a^2) + L^2 \Delta_r^2 - r^4 = 0.
	\label{54}
	\end{equation}
	From this equation, the energy $E$ and angular momentum $L$ of the particle can be expressed as
	\begin{equation}
	E = \frac{a (r^2 + a^2) L \pm r^2}{(r^2 + a^2)^2},
	\label{55}
	\end{equation}
	\begin{equation}
	L = \frac{a (r^2 + a^2) E \pm r^2}{\Delta_r}.
	\label{55}
	\end{equation}
	
	To analyze negative energy states ($E < 0$), both $E$ and $L$ must be negative. This interplay depends on the spin parameter $a$, angular momentum $L$, and radial position $r$. For $L < 0$, the spin parameter $a$ must be positive, ensuring counter-rotating orbits. Additionally, the condition for negative angular momentum can be written as
	\begin{equation}
	L < -\frac{r^2}{a (r^2 + a^2)}.
	\label{56}
	\end{equation}
	
	Here, $a > 0$ ensures the necessary coupling between the spin of the black hole, the particle's angular momentum, and its radial location. These conditions highlight how counter-rotating orbits and electromagnetic effects can produce negative energy states, particularly in spacetimes with high spin or significant charge.
	
	To calculate the Penrose mechanism efficiency and the energy of particles in the context of the KNAdS metric with quintessence matter, we begin by determining the energy of the incoming particle. This energy can be derived from the components of the metric by evaluating the specific energy of the particle at a specific radial position $r$. 
	
	For a particle moving in the equatorial plane ( $\theta = \pi/2$) of the KNAdS BH, the energy $E_{\text{in}}$is given by,
	\begin{equation}
	E_{\text{initial}} = - g_{tt} u^t - g_{t\phi} u^\phi
	\label{047}
	\end{equation}
	where $u^\mu = \frac{dx^\mu}{d\tau}$  is the four-velocity of the particle, and  $\tau$  is the proper time. For a particle moving on a geodesic orbit, the four-velocity components  $u^t$  and  $u^\phi$  can be calculated from the conserved quantities that are energy $E$ and angular momentum  $L$. We can use the relations for the energy  $E$ and angular momentum  $L$ of a particle orbiting a BH,
	\begin{equation}
	E = - g_{tt} \frac{dt}{d\tau} - g_{t\phi} \frac{d\phi}{d\tau} 
	\end{equation}
	\begin{equation}
	L = g_{t\phi} \frac{dt}{d\tau} + g_{\phi\phi} \frac{d\phi}{d\tau}
	\end{equation}
	These two quantities are conserved for a particle orbiting the BH in the equatorial plane. 
	
	To determine the energy extraction through the Penrose process, the first step is to locate the ergoregion, situated between the outer event horizon ($r_+$) and the static limit, where $g_{tt} > 0$. Within this region, a particle undergoes a splitting process: one fragment is absorbed by the black hole, leading to a reduction in its angular momentum, while the other fragment escapes to infinity with an energy greater than that of the original particle, thereby facilitating the extraction of energy from the BH.
	\begin{equation}
	\Delta E = E_{\text{final}} - E_{\text{initial}},
	\end{equation}
	where $E_{\text{final}} > E_{\text{initial}}$. For a particle in this spacetime, the energy is associated with the conserved quantity corresponding to the timelike Killing vector $\xi^\mu = \partial_t$. The energy of a particle can be expressed as,
	\begin{equation}
	E = -p_\mu \xi^\mu = -p_t.
	\end{equation}
	The components of the 4-momentum $p_\mu$ are related to the motion of the particle through the metric, so we need to compute the $t$-component of the 4-momentum,
	\begin{align}
	p_t = \frac{\Sigma^2}{\Delta_r} \dot{r}^2 +& \frac{\Sigma^2}{\Delta_\theta} \dot{\theta}^2 
	+ \frac{\Delta_\theta \sin^2 \theta}{\Sigma^2} \left(a \dot{t} - (r^2 - a^2) \dot{\phi}\right)^2 \notag \\
	&\quad - \frac{\Delta_r}{\Sigma^2} \left( \dot{t} - a \sin^2 \theta \dot{\phi} \right)^2.
	\end{align}
	At infinity, for a particle escaping the BH, the radial component becomes zero (\(\dot{r} \to 0\)), and the energy of the particle is given by,
	\begin{equation}
	E_{\infty} = -p_t \Big|_{\text{escape}} = M + \frac{Q^2}{2\ell^2} + \frac{3}{8 \pi \ell^2}.
	\end{equation}
	This represents the total energy of the particle as it reaches infinity. The energy is higher than the initial value because the particle escapes with excess energy, leading to a loss of mass or angular momentum by the BH.  The energy of the particle before the split is given by
	\begin{equation}
	E_{\text{initial}} = -p_t \Big|_{\text{inside ergoregion}}.
	\end{equation}
	Inside the ergoregion, $g_{tt} > 0$, and we need the particle's energy for a general geodesic motion, which involves solving for the $t$-component of the energy using the metric's conserved quantities. Since the initial energy depends on the specific trajectory of the particle inside the ergoregion, we have,
	\begin{equation}
	\Delta E = \left( M + \frac{Q^2}{2\ell^2} + \frac{3}{8 \pi \ell^2} \right) - E_{\text{initial}}.
	\end{equation}
	The exact value of \(E_{\text{initial}}\) depends on the particle's motion and position inside the ergoregion. In a simplified case, if we assume that the initial energy is purely due to the black hole’s gravitational energy and angular momentum, the energy extraction could be expressed in terms of the BH's mass \(M\), charge \(Q\), and cosmological constant \(\ell\).
	
	The 4-momentum components are related to the motion of the particle. Specifically, for a particle moving in the equatorial plane, the energy is given by (\ref{047}) and we have the following relations for the components of the 4-velocity.
	$\dot{t} = \frac{E_{\text{initial}}}{g_{tt}}$, $\dot{\phi} = L / g_{t\phi}$, where $L$ is the conserved angular momentum per unit mass.
	
	To maintain simplicity and ensure physical justification, we consider a stationary radius close to the event horizon but still within the ergoregion. At this location, the energy can be approximated as,
	\begin{equation}
	E_{\text{initial}} \approx \left( 1 - \frac{2M}{r_+} + \frac{a^2}{r_+^2} + \frac{r_+^2}{\ell^2} \right).
	\end{equation}
	This formula reflects the energy of a particle that is near the event horizon and is being pulled into the black hole, where the gravitational potential is very high.  The difference between the energy of the escaping particle and the energy of the particle inside the ergoregion gives us the energy extracted. Since the particle escapes with excess energy, we compute the total extracted energy as,
	\begin{equation}
	\Delta E = E_{\infty} - E_{\text{initial}}.
	\end{equation}
	Thus, the energy extracted via the Penrose process is,
	\begin{equation}
	\Delta E = \left( M + \frac{Q^2}{2\ell^2} + \frac{3}{8 \pi \ell^2} \right) - \left( 1 - \frac{2M}{r_+} + \frac{a^2}{r_+^2} + \frac{r_+^2}{\ell^2} \right),
	\end{equation}
	where \(r_+\) is the location of the event horizon, which is determined by solving \(\Delta_r = 0\) for \(r_+\).
	\subsection {Efficiency of the Penrose Process}
	The efficiency of the Penrose process is defined as the fraction of the initial energy that is extracted. It can be written as
	\begin{equation}
	\eta = \frac{\Delta E}{E_{\text{initial}}}.
	\end{equation}
	Substituting the expressions for \(\Delta E\) and \(E_{\text{initial}}\), we get the simplified expression for $\eta$,
	\begin{equation}
	\eta = \frac{M\left( 1 + \frac{2}{r_+} \right) + \frac{Q^2}{2\ell^2} + \frac{3}{8 \pi \ell^2} - 1 - \frac{a^2}{r_+^2} - \frac{r_+^2}{\ell^2}}{1 - \frac{2M}{r_+} + \frac{a^2}{r_+^2} + \frac{r_+^2}{\ell^2}}.
	\end{equation}
	For large values of $r_+$, the term involving $r_+^2/\ell^2$ becomes significant, especially in the case of AdS black holes. In the case of a maximally rotating black hole ($a = r_+$), the efficiency can approach its maximum value. In this case, the energy extraction becomes more efficient because the BH spin contributes to a greater difference between the initial and final energy. 
	
	The efficiency (\(\eta\)) of the Penrose mechanism for the KNAdS BH is strongly influenced by the spin parameter (\(a/M\)), the charge-to-mass ratio (\(Q/M\)), and the quintessence parameter (\(\alpha\)). As shown in Fig. 8(a), increasing the spin parameter initially enhances the efficiency due to stronger frame-dragging effects caused by higher rotation. However, for larger values of the quintessence parameter (\(\alpha\)), the efficiency decreases, as quintessence energy exerts a repulsive effect that stretches the ergosphere outward, reducing the interaction strength between particles and the black hole's rotational energy. This trend is further illustrated in Fig. 8(b), where efficiency diminishes as \(\alpha\) increases for different spin values, showing that the repulsive effects of quintessence dominate the rotational dynamics of the BH.

	The influence of the charge-to-mass ratio (\(Q/M\)) and radial distance (\(r/M\)) is further explored in Fig. 8(c)-(e). In Fig. 8(c), a higher \(Q/M\) significantly reduces efficiency, especially for larger \(\alpha\), as higher charge suppresses frame-dragging effects and diminishes the rotational energy available for extraction. Similarly, Fig. 8(d) shows that efficiency peaks near the ergosphere and rapidly declines outward, reflecting the localized nature of energy extraction. Fig. 8(e) highlights the combined effects of \(Q/M\) and \(\alpha\), with higher charge and quintessence energy drastically reducing efficiency. This behavior underscores the complex interplay of parameters, where spin enhances efficiency, while higher charge and quintessence energy act as limiting factors by weakening the frame-dragging effects and stretching the ergosphere.

	The table 1. presents the efficiency (\(\eta\)) of the Penrose mechanism for different  parameters. It shows that higher spin values (\(a/M\)) and lower radial distances enhance efficiency, while increasing \(\alpha\) and charge (\(Q/M\)) generally reduce energy extraction. This highlights the influence of  parameters on the efficiency of the Penrose process, suggesting that rapidly rotating black holes with minimal quintessence effects are the most favorable for extracting energy.

	\begin{table}[h!]
		\centering
		\begin{tabular}{|c|c|c|c|c|}
			\hline
			$a/M$ & $\alpha$ & $Q/M$ & $r/M$ & Efficiency ($\eta$) (\%) \\  
			\hline
			0.5 & 0.1 & 1.0 & 1.5 & 25\% \\  
			1.0 & 0.1 & 0.5 & 1.5 & 35\% \\  
			1.5 & 0.3 & 1.5 & 2.0 & 10\% \\  
			0.5 & 0.5 & 1.0 & 2.5 & 5\% \\  
			1.0 & 0.5 & 0.5 & 1.5 & 15\% \\  
			\hline
		\end{tabular}
		\caption{Table for Efficiency ($\eta$) of the Penrose Mechanism in Percentages.}
	\end{table}
	\begin{figure*}[ht]
		\centering
		\begin{minipage}{0.38\textwidth}
			\centering
			\includegraphics[width=1\linewidth]{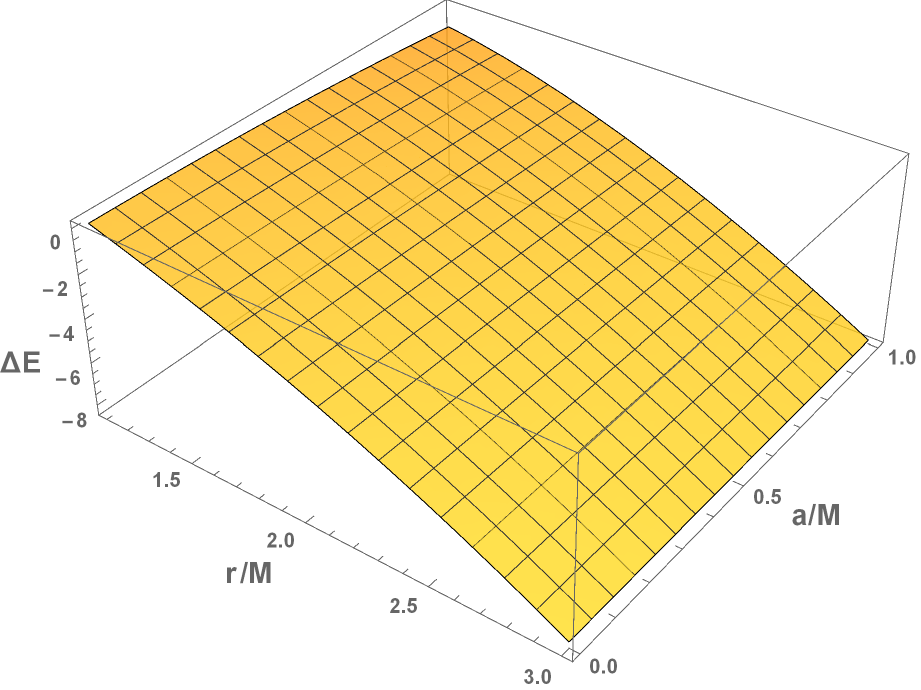}
			{Fig. 7(a)}
		\end{minipage}
		\hspace{1.8cm}
		\begin{minipage}{0.38\textwidth}
			\centering
			\includegraphics[width=1\linewidth]{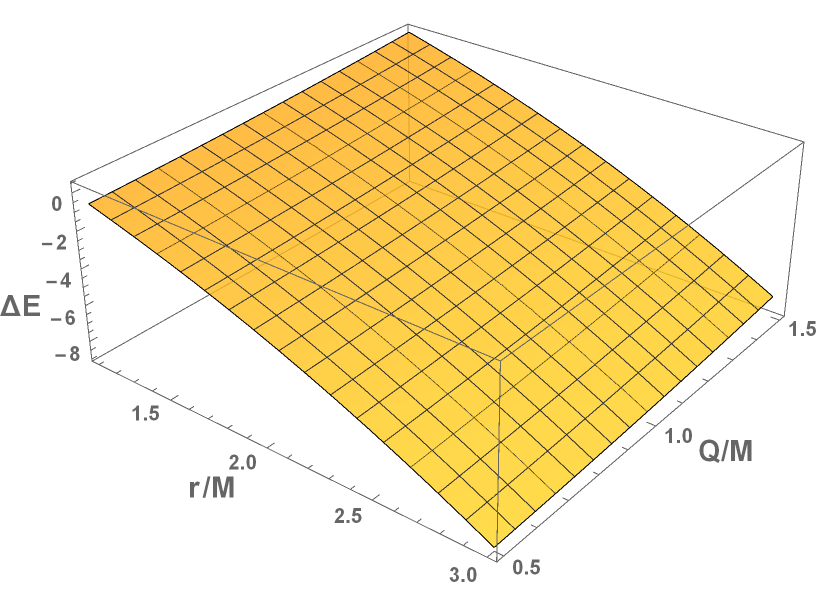}
			{Fig.7(b)}
		\end{minipage}%
		~
		
		\vspace{.7cm}
		\begin{minipage}{0.38\textwidth}
			\centering
			\includegraphics[width=1\linewidth]{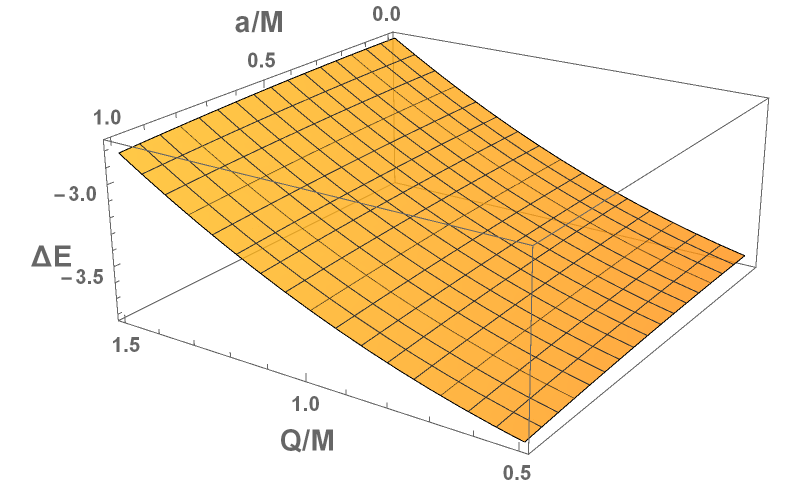}
			{Fig. 7(c)}
		\end{minipage}%
		\captionsetup{justification=raggedright,singlelinecheck=false}
		\caption{Energy extracted via the Penrose process (\(\Delta E\)) for a KNAdS BH with quintessence energy, plotted as a function of:(a) Spin parameter (\(a/M\)) and radial coordinate (\(r/M\)): Energy extraction is maximized at intermediate radii near the ergosphere and decreases with increasing spin (\(a/M\)) due to weaker frame-dragging effects at larger radii.  (b) Charge-to-mass ratio (\(Q/M\)) and radial coordinate (\(r/M\)): The extracted energy decreases as \(Q/M\) increases because higher black hole charge reduces the available rotational energy in the ergosphere, with maximum extraction occurring near smaller radii.  (c) Spin parameter (\(a/M\)) and charge-to-mass ratio (\(Q/M\)): Energy extraction efficiency decreases with higher \(Q/M\) and \(a/M\), as the combination of high charge and spin diminishes the available rotational energy.}
	\end{figure*}
	\begin{figure*}[t!]
		\centering
		\begin{minipage}[t]{0.38\textwidth}
			\centering
			\includegraphics[width=1\linewidth]{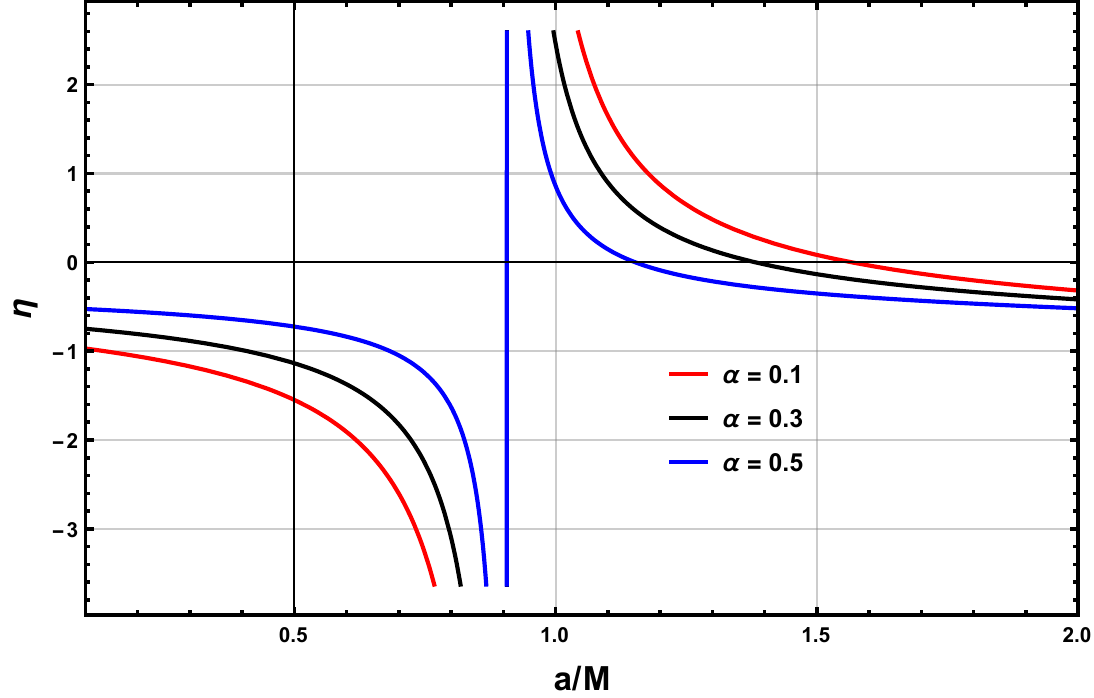}
			{Fig. 8(a)}
		\end{minipage}
		\hspace{1.8cm}
		\begin{minipage}[t]{0.38\textwidth}
			\centering
			\includegraphics[width=1\linewidth]{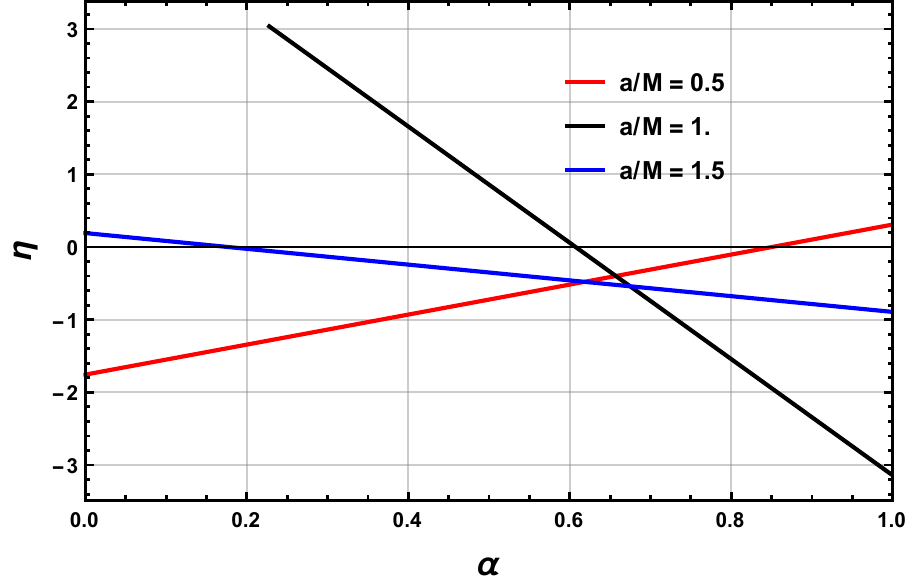}
			{Fig.8(b)}
		\end{minipage}%
		~
		
		\vspace{.7cm}
		\begin{minipage}[t]{0.38\textwidth}
			\centering
			\includegraphics[width=1\linewidth]{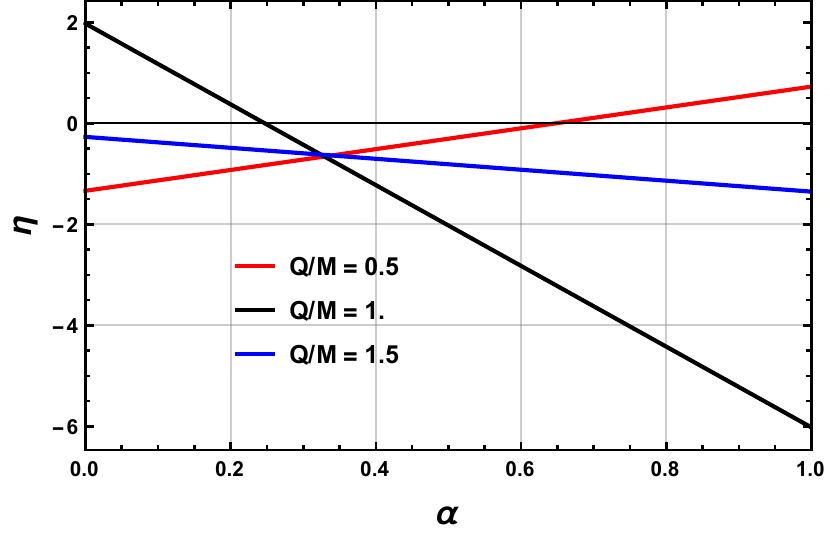}
			{Fig. 8(c)}
		\end{minipage}%
		\hspace{1.8cm}
		\begin{minipage}[t]{0.38\textwidth}
			\centering
			\includegraphics[width=1\linewidth]{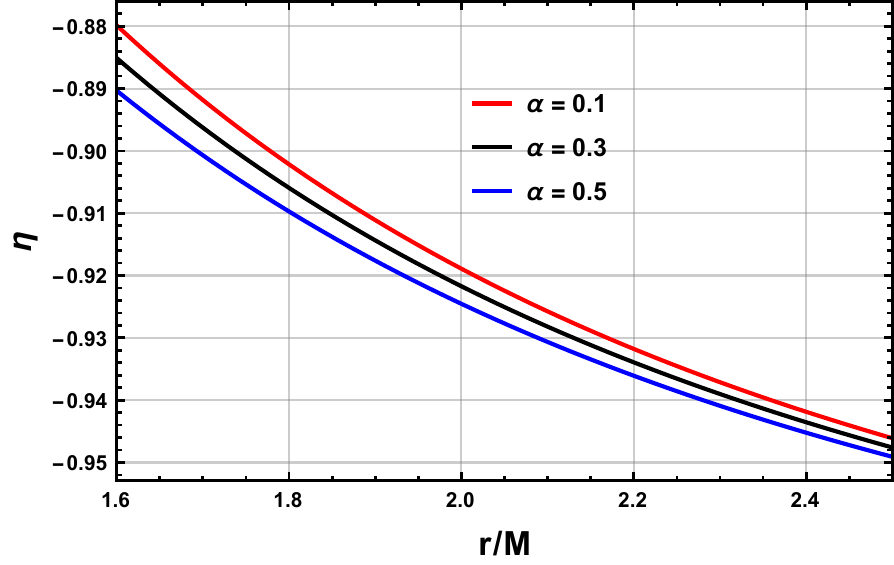}
			{Fig.8(d)}
		\end{minipage}%
		~
		
		\vspace{.7cm}
		\begin{minipage}[t]{0.38\textwidth}
			\centering
			\includegraphics[width=1\linewidth]{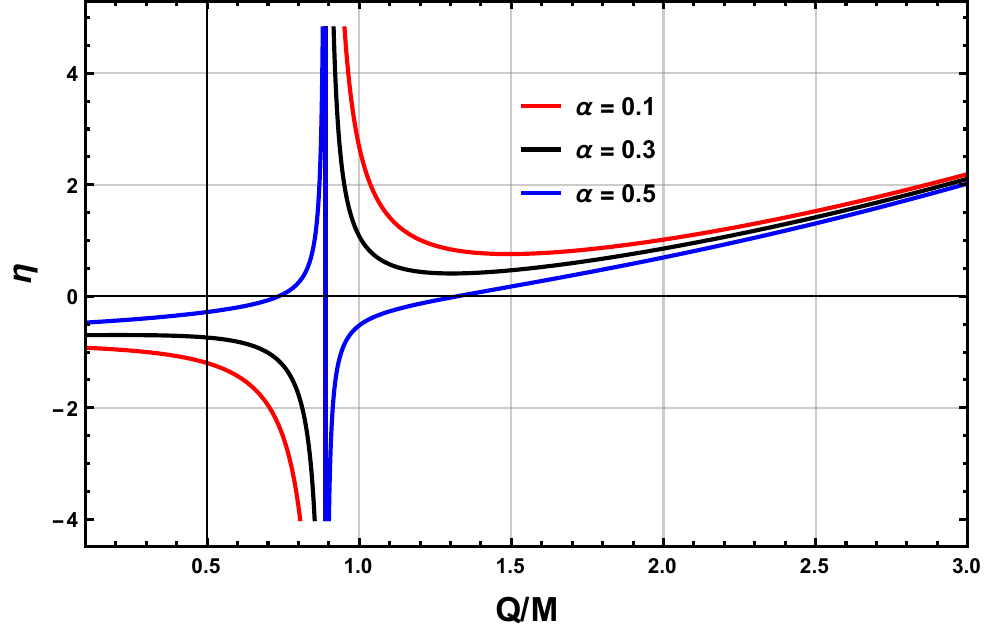}
			{Fig.8(e)}
		\end{minipage}%
		\captionsetup{justification=raggedright,singlelinecheck=false}
		\caption{Efficiency (\(\eta\)) of the Penrose mechanism for the Kerr-Newman-AdS black hole, analyzed with respect to various parameters: (a) Efficiency \(\eta\) as a function of spin parameter (\(a/M\)) for different quintessence parameter values (\(\alpha = 0.1, 0.3, 0.5\)). Higher spin enhances efficiency initially but becomes limited due to the repulsive effect of quintessence energy as \(\alpha\) increases. (b) Efficiency \(\eta\) versus quintessence parameter \(\alpha\) for different spin parameters (\(a/M = 0.5, 1, 1.5\)). The increase in \(\alpha\) reduces the energy extraction efficiency due to the outward expansion of the ergosphere. (c) Efficiency \(\eta\) as a function of \(\alpha\) for varying charge-to-mass ratios (\(Q/M = 0.5, 1, 1.5\)). Higher \(Q/M\) values reduce efficiency, with a steep decline for larger \(\alpha\), indicating diminished frame-dragging effects. (d) Efficiency \(\eta\) as a function of radial distance (\(r/M\)) for different \(\alpha\) values. Efficiency peaks near the ergosphere and declines outward as the energy available for extraction reduces. (e) Efficiency \(\eta\) as a function of \(Q/M\) for varying \(\alpha\). A higher charge significantly suppresses efficiency, especially for larger \(\alpha\), due to reduced rotational energy in the system.}
	\end{figure*}
	
	\subsection{Irreducible Mass}
	When extracting energy via the Penrose process in a Kerr BH, there exists an upper limit to the amount of energy that can be extracted. In particular, it is not possible to extract energy until the BH loses all of its mass. Consequently, there is an irreducible energy associated with the bH, which, in turn, corresponds to a mass known as the \textit{irreducible mass} $M_{\text{irr}}$. This concept was first introduced by Christodoulou \cite{christodoulou1970reversible}, who derived the expression for $M_{\text{irr}}$ by considering the maximum efficiency of the Penrose process for a test particle in the equatorial plane. 
	
	A closely related result, formulated directly in terms of the irreducible energy, can be obtained within the framework of the \textit{Teleparallel Equivalent of General Relativity} (TEGR), a dynamically equivalent formulation of GR that allows for a well-defined energy-momentum tensor for the gravitational field. By integrating this energy-momentum tensor over the event horizon, one can evaluate the total gravitational energy contained within it, which corresponds to the irreducible energy of the black hole \cite{maluf1996gravitational}.
	
	An alternative approach to determining the irreducible mass follows from the singularity theorems of Penrose, which state that the area of the event horizon cannot decrease. Thus, as energy is extracted from the black hole, the stationary surface approaches the event horizon, and when the two surfaces coincide, no further energy can be extracted.
	
	There exists an intrinsic relationship between the area of the BH's horizon and its irreducible mass. Consider, for instance, a Schwarzschild black hole, where the mass parameter is identical to its irreducible mass since the stationary surface already coincides with the event horizon. For a Schwarzschild BH, the event horizon area is given by  
	\begin{equation}
	A=4\pi R^{2}\,,
	\end{equation}  
	where \(R = 2m\) is the radius of the event horizon. Since in this case \(m = M_{\text{irr}}\), we obtain  
	\begin{equation}\label{64}
	A=16\pi M_{\text{irr}}^{2}\,,
	\end{equation}  
	i.e., the event horizon area is proportional to the square of the irreducible mass.
	
	For the metric associated with the line element (\ref{01}), the area of a two-dimensional surface spanned by the coordinates \((\theta, \phi)\) is given by  
	\begin{equation}
	A=\oint |g_{22}g_{33}|^{1/2}d\theta d\phi\,,
	\end{equation}  
	where \( g_{22}g_{33}=\frac{1}{\Delta_{\theta}\Xi} \Big( \Delta_{\theta}(r^{2}+a^{2})^{2}-\Delta_{r}a^{2}\sin^{2}{\theta}\Big) \). At the event horizon \(r = r_+\), where \(\Delta_r = 0\), we obtain  
	\begin{align}
	A&=2\pi\int_{0}^{\pi}\frac{(r_{+}^{2}+a^{2})}{1-a^{2}/l^{2}}\sin{\theta}\,d\theta\nonumber\\
	&=4\pi \frac{r_{+}^{2}+a^{2}}{1-a^{2}/l^{2}}\,.\label{66}
	\end{align}  
	Comparing (\ref{66}) with (\ref{64}), we find the irreducible mass of the KNAdS BH as  
	\begin{equation}
	M_{\text{irr}}=\frac{1}{2}\frac{\sqrt{r_{+}^{2}+a^{2}}}{\sqrt{1-a^{2}/l^{2}}}\,.
	\end{equation}  
	For consistency, we verify that taking the limit \( l \to \infty \) recovers the irreducible mass of a Kerr bh \(M_{\text{irr}}=\frac{1}{2}\sqrt{r_{+}^{2}+a^{2}}\). By comparing these expressions, we observe that the presence of the cosmological parameter increases the irreducible mass. Since the maximum amount of energy that can be extracted is given by \( M - M_{\text{irr}} \), an increase in \( M_{\text{irr}} \) implies that less energy can be extracted. The quintessence parameter \( \alpha \), while appearing in \( r_+ \), does not directly modify the irreducible mass but influences it indirectly by altering the radius of the event horizon \( r_+ \). 
	\section{Conclusion and Discussion}
	In this study, we have examined the structural and dynamical properties of KNAdS BHs within the context of quintessence energy. We discussed the influence of quintessence and the rotation parameter on key regions such as photon regions, ergospheres, horizons, and static limits. Our analysis sheds light on how parameters like spin (\(a\)), charge-to-mass ratio (\(Q/M\)), quintessence parameter (\(\alpha\)), and the equation of state parameter (\(\omega\)) impact the spacetime geometry and the behavior of matter around these BH.
	
	Our findings reveal that quintessence significantly alters the structure of the KNAdS BH. The introduction of quintessence modifies the ergosphere, static limit, photon region, and event horizons, leading to distinct geometric changes depending on the values of the parameters.
	
	We found that the static limit is strongly influenced by spin and quintessence, with lower spins leading to nearly spherical static limits and higher spins causing elongation along the equatorial plane. The photon region  becomes more asymmetric with increasing spin and quintessence, reflecting the impact of these parameters on null geodesics. As the quintessence parameter and spin increase, the ergosphere elongates, and the static limit shifts outward, creating larger regions where particles cannot remain stationary. In extreme cases, the ergosphere becomes highly deformed, and the photon region may split or become disconnected. These results underscore the significant sensitivity of the BH geometry to the interplay of spin and quintessence. The behavior is shown in Fig. 2.
	
	The energy $E$ and angular momentum $L$ of test particles around the KNAdS BH were analyzed in the presence of quintessence. Fig. 3 shows the energy profiles for prograde and retrograde orbits, revealing that prograde motion requires less energy as the spin parameter increases, while retrograde motion requires higher energy. The BH mass also influences the energy, with higher masses lowering the energy requirement. Fig. 4 illustrates the angular momentum profiles, showing that prograde orbits exhibit lower specific angular momentum $L$ compared to retrograde orbits. This is due to the frame-dragging effect, which benefits prograde motion but opposes retrograde motion.
	
	The stability of particle orbits was analyzed using the effective force and Lyapunov exponents. Fig. 5 shows the effective force acting on particles, revealing that higher energy and angular momentum  reduce the steepness of the force near smaller radii. Quintessence energy modifies the effective potential, shifting these effects outward. Fig. 6 presents the Lyapunov exponents, which quantify orbital stability. Higher energy and angular momentum extend the region of instability (positive 
	$\lambda$) outward, indicating that particles can explore chaotic orbits farther from the BH. Quintessence energy further enhances this outward shift due to its repulsive nature.
	
	We examined the Penrose process for energy extraction from the ergoregion of the KNAdS BH. Fig. 7 shows the energy extracted ($\Delta E$ ) as a function of spin ($a/M$), charge-to-mass ratio ($Q/M$), and radial distance (
	$r/M$). The results indicate that energy extraction is maximized near the ergosphere and decreases with increasing spin and charge. Fig. 8 provides the efficiency ($\eta$) of the Penrose process, showing that efficiency peaks at high spin and low quintessence parameter ($\alpha$). Higher charge and quintessence energy reduce efficiency by weakening the frame-dragging effects and stretching the ergosphere.
	
	We have derived the expression for the irreducible mass of a KNAdS BH and analyzed its dependence on the cosmological and quintessence parameters. We found that the presence of a negative cosmological constant increases the irreducible mass, thereby reducing the maximum extractable energy. While the quintessence parameter does not explicitly appear in the expression for \( M_{\text{irr}} \), it influences the result indirectly by modifying the event horizon radius.   
	
	This study opens several avenues for future research, including the exploration of observational signatures of KNAdS BHs with quintessence and their implications for high-energy astrophysics. Additionally, further investigations could focus on the interaction of quintessential energy with other types of BHs and the potential observational consequences of these interactions.
	\bibliographystyle{unsrt}
	\bibliography{bibitex}
\end{document}